\documentclass[preprint,12pt]{elsarticle}
\usepackage{amssymb}
\usepackage{amsmath}
\usepackage{xcolor}
\usepackage{amsthm}
\usepackage{lineno}
\usepackage{float}
\usepackage{subcaption}
\usepackage[shortcuts]{extdash}
\usepackage[commentmarkup=uwave]{changes}
\definechangesauthor[name={Stefano Palagi}, color=blue]{SP}
\definechangesauthor[name={Jyoti Sharma}, color=red]{JS}
\usepackage[version=4]{mhchem}
\usepackage[separate-uncertainty = true,multi-part-units=single]{siunitx}
\usepackage{geometry}
\usepackage[hyphens]{url}
\journal{ACS Nano}

\AtBeginDocument{
  \def\thefnote{\myfnsymbol{fnote}}}
\makeatletter
\def\myfnsymbol#1{\expandafter\@myfnsymbol\csname c@#1\endcsname}
\def\@myfnsymbol#1{\ifcase #1\or $\dagger$\or $\dagger\dagger$\else \@ctrerr\fi}
\def\fntext[#1]#2{\g@addto@macro\@fnotes{
   \refstepcounter{fnote}\elsLabel{#1}
   \def\thefootnote{\thefnote}
   \global\setcounter{footnote}{\c@fnote}
   \footnotetext{#2}}}
\makeatother

\begin{document}
 
\begin{frontmatter}
\title{Patterned Substrates Unlock Self-Electrophoretic Phenomenon in Active Janus Microswimmers.}

\author[SSSA]{Jyoti Sharma\fnref{equal}\corref{cor1}}\ead{jyotisharmaphy@gmail.com}
\author[SSSA,WUR]{Yashpal Singh Brar\fnref{equal}}
\author[SSSA]{Omar Tricinci}
\author[IIT]{Paola Parlanti}
\author[IIT]{Mauro Gemmi}
\author[SSSA]{Stefano Palagi}
\fntext[equal]{Jyoti Sharma and Yashpal Singh Brar contributed equally to this work.}

\affiliation[SSSA]{organization={Istituto di BioRobotica-Scuola Superiore Sant'Anna},
            addressline={Viale Rinaldo Piaggio, 34}, city={Pontedera},
            postcode={56025}, state={Pisa}, country={Italy}}

\affiliation[IIT]{organization={Center for Materials Interfaces, Electron Crystallography, Italian Institute of Technology},
            addressline={Viale Rinaldo Piaggio, 34}, city={Pontedera},
            postcode={56025}, state={Pisa}, country={Italy}}   

\affiliation[WUR]{organization={Physical Chemistry and Soft Matter Group\\Wageningen University \& Research},
            addressline={Helix, Stippeneng 4}, postcode={6708 WE},
            city={Wageningen}, state={Gelderland}, country={The Netherlands}}
            
\cortext[cor1]{Corresponding author.}

\begin{abstract}
Inert colloids half-coated with platinum (\ce{Pt}) are a standard model of chemically powered active particles, yet the microscopic origins of their propulsion in hydrogen peroxide (\ce{H2O2})remain difficult to dissect experimentally. Whereas self-diffusiophoresis was the prevailing theory, self-electrophoresis has been more recently suggested as the main mechanism of propulsion. According to the latter mechanism, the pole-to-equator \ce{Pt}-thickness gradient produced by directional metal deposition is sufficient to create anodic and cathodic regions on the metal cap and thereby generate an electric field sustained by \ce{H2O2} decomposition. Enhancing self-propulsion performance of such particles thus requires precise control over the \ce{Pt} thickness distribution, which is currently not achievable with standard methods (e.g. evaporation or sputtering). Here, we propose a method to fabricate Janus active particles by assembling silica microspheres on patterned substrates containing spherical grooves whose depth and spacing set the degree of particle coating while simultaneously suppressing proximity-led defects (\ce{Pt} bridges). The resulting particles exhibit a tunable platinum-thickness contrast, as verified by Focused-Ion-Beam cross-sections. In 2.5\,\% \ce{H2O2}, our results suggest that this control can significantly increase propulsion efficiency, while providing evidence indirectly supporting the hypothesis that self-electrophoresis is the dominant mechanism. These results demonstrate that our patterned-substrate route can enhance control over the catalyst deposition and enable novel Janus morphologies, allowing for more precise engineering of active colloids.
\end{abstract}

 
\clearpage
\end{frontmatter}

\section{Introduction}\label{intro}
Autonomous Janus microswimmers are a frontier in physics for studying non-equilibrium active matter systems and hold strong potential for targeted drug delivery, microscale cargo transport, and environmental sensing~\cite{gompper_2025_2025, bechinger_active_2016, fengAdvancesChemicallyPowered2023, sanchez_chemically_2015, gao_synthetic_2014, peng_micronanomotors_2017, zhang_micronano-motors_2025, juTechnologyRoadmapMicro2025}. Janus microswimmers are a prominent class of self-propelled colloids that possess two chemically distinct hemispheres, typically one catalytic and one inert. These particles are conventionally prepared by assembling a colloidal monolayer on a flat substrate (e.g.\ by drop-casting or by a Langmuir-Blodgett technique), followed by deposition of a catalytic material (e.g.\ by sputtering or evaporation)~\cite{loveFabricationWettingProperties2002, howseSelfMotileColloidalParticles2007, gibbsCatalyticNanomotorsFabrication2011, paxton_catalytic_2004}. Among the wide range of Janus colloids reported to date, platinum (\ce{Pt})-based Janus particles remain the most widely explored, due to their robust autonomous motion in dilute hydrogen peroxide (\ce{H2O2}) solutions~\cite{howseSelfMotileColloidalParticles2007, ebbens_pursuit_2010}. The \ce{Pt} hemisphere acts as a catalyst for a key reaction, i.e.\ the decomposition of \ce{H2O2} into water and oxygen, which is responsible for the motion.

The microscopic origin of this motion has been debated for more than a decade~\cite{howseSelfMotileColloidalParticles2007, wang_open_2023}. 
Early interpretations invoked oxygen-bubble recoil or neutral self-diffusiophoresis arising from concentration gradients of \ce{H2O2} and \ce{O2} around the particle as a plausible propulsion mechanism~\cite{gibbsAutonomouslyMotileCatalytic2009, golestanianPropulsionMolecularMachine2005}. However, subsequent observations that added electrolytes strongly suppress swimming speed, and that direction can reverse with surface-charge modification, pointed to the involvement of charged intermediates and a self-generated electric field—i.e.\ self-electrophoresis~\cite{brownIonicEffectsSelfpropelled2014a, wuZetaPotentialDependent2017, ebbens_electrokinetic_2014}.

According to the self-electrophoresis hypothesis, the particle motion is driven by a local electric field generated by the asymmetric electrocatalytic decomposition of \ce{H2O2} on the \ce{Pt} hemisphere~\cite{kuronUnderstandingSelfElectrophoreticPropulsion2018}. Although there is strong support for self-electrophoresis as the dominant mechanism for catalytic Janus microswimmers~\cite{brownIonicEffectsSelfpropelled2014a, ebbens_electrokinetic_2014}, truly direct microscopic evidence of the self-generated electric field or ionic current remains challenging. Nonetheless, indirect experimental evidence commonly used to support self-electrophoresis includes: zeta-potential-dependent velocity of Janus microswimmers~\cite{wuZetaPotentialDependent2017}, flow field around the particle~\cite{heckelJanusGeometryCharacterization2022}, and \ce{H2O2} consumption-driven reaction-rate measurements \cite{ebbens_electrokinetic_2014, brownIonicEffectsSelfpropelled2014a}. Lyu {\it et al.}\ recently provided another novel approach, based on a \ce{Pt} thickness difference, to further support claims of self-electrophoresis~\cite{lyu2021active}.

A key source of catalytic asymmetry is the inherent thickness gradient produced by directional metal deposition via sputtering or evaporation: the coating is thicker at the pole than near the equator~\cite{rashidi_local_2018}, as illustrated in Figure~\ref{fig:reaction_schematic}. Because the reaction rate on \ce{Pt} depends on film thickness in the few-nanometer regime~\cite{ebbens_electrokinetic_2014, ibrahimMultiplePhoreticMechanisms2017}, this gradient can itself generate the anodic/cathodic differentiation required for self-electrophoresis. The thinner equator is believed to function as the anode (oxidizing \ce{H2O2} and releasing protons and electrons), the thicker pole functions as the cathode (reducing \ce{H2O2} to water and consuming electrons)~\cite{lyu2021active}. The resulting local electric field propels the particle toward the inert hemisphere. If this thickness-contrast picture is correct, systematically increasing the pole-to-equator \ce{Pt} contrast at fixed fuel concentration should increase propulsion speed. Isolating and systematically varying this nanoscale thickness contrast, however, has proven difficult using conventional fabrication methods.

Here, we introduce a fabrication strategy that uses geometrically defined embedding of silica microspheres in patterned substrates to vary the pole-to-equator \ce{Pt} thickness contrast after sputtering, while suppressing proximity-caused imperfections. Focused-ion-beam cross-sections confirm that shallower embedding yields a more pronounced thickness contrast. Particles prepared in this way swim faster in dilute \ce{H2O2} as the designed contrast increases. The correlation between thickness asymmetry and speed provides experimental support for thickness-gradient-driven self-electrophoresis, and shows that control of the \ce{Pt} profile is a direct handle on propulsion efficiency at fixed fuel concentration.

\section{Results}
\subsection{\textbf{Spaced monolayers enable precise and controllable Janus asymmetry}} 

\begin{figure}[htbp]
    \centering
    \begin{subfigure}[c]{0.49\textwidth}
        \centering
        \includegraphics[width=\textwidth]{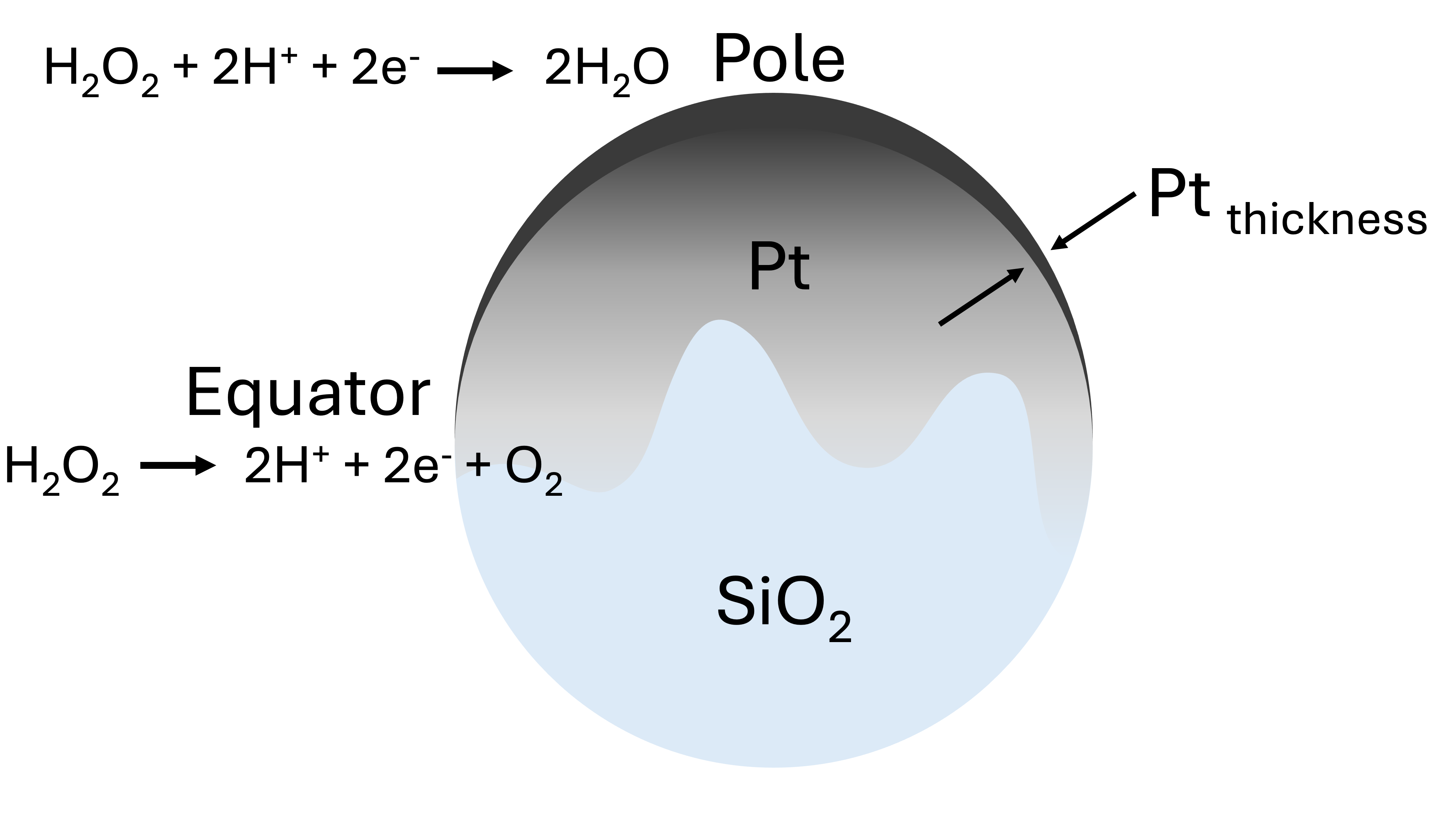}
        \caption{}\label{fig:reaction_schematic}
    \end{subfigure}
    \hfill
    \begin{subfigure}[c]{0.49\textwidth}
        \centering
        \includegraphics[width=\textwidth]{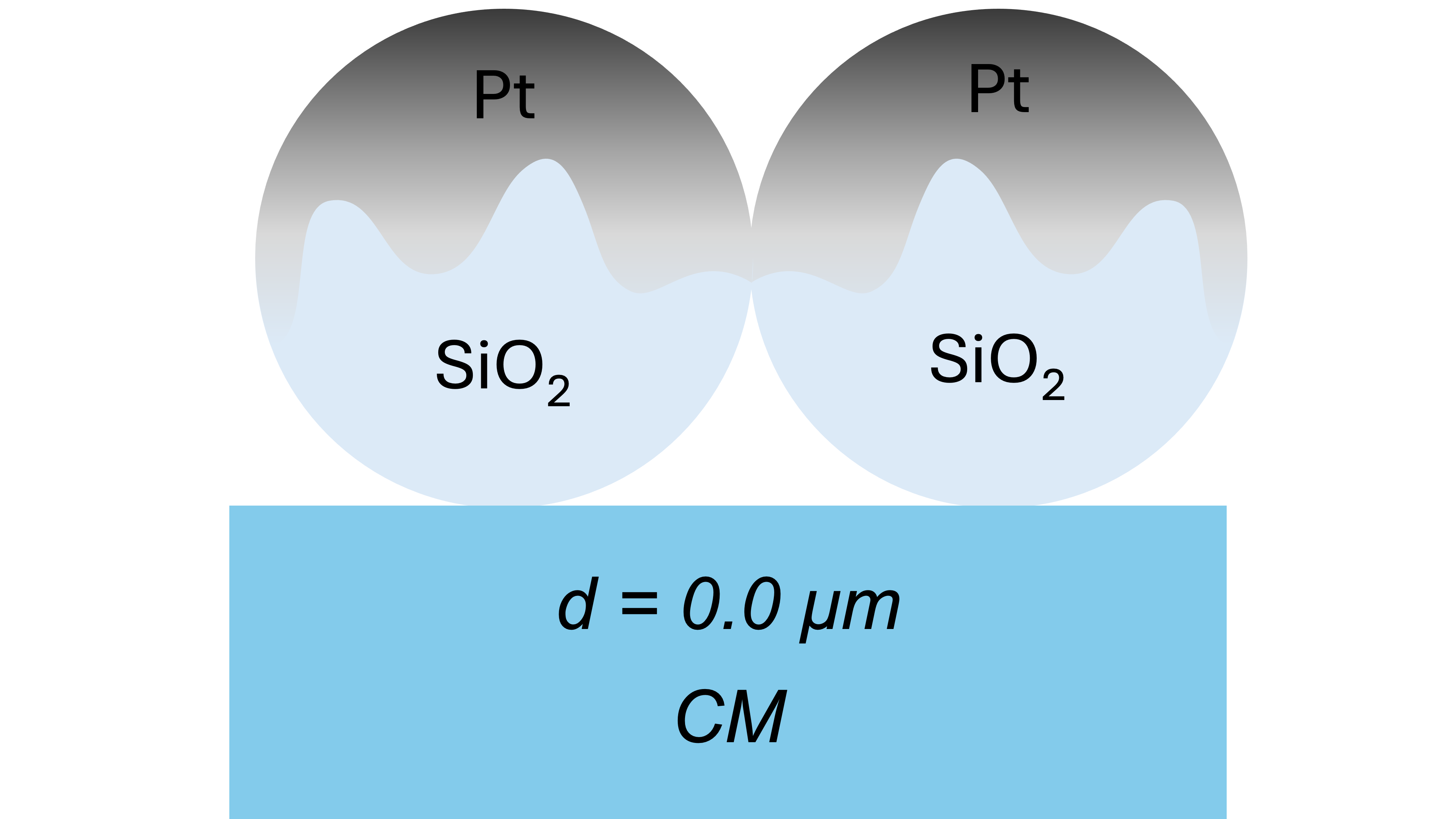}
        \caption{}\label{fig:schematic_control}
    \end{subfigure}

    \vspace{1.0em} 
    
    \begin{subfigure}[b]{0.32\textwidth}
        \centering
        \includegraphics[width=\textwidth]{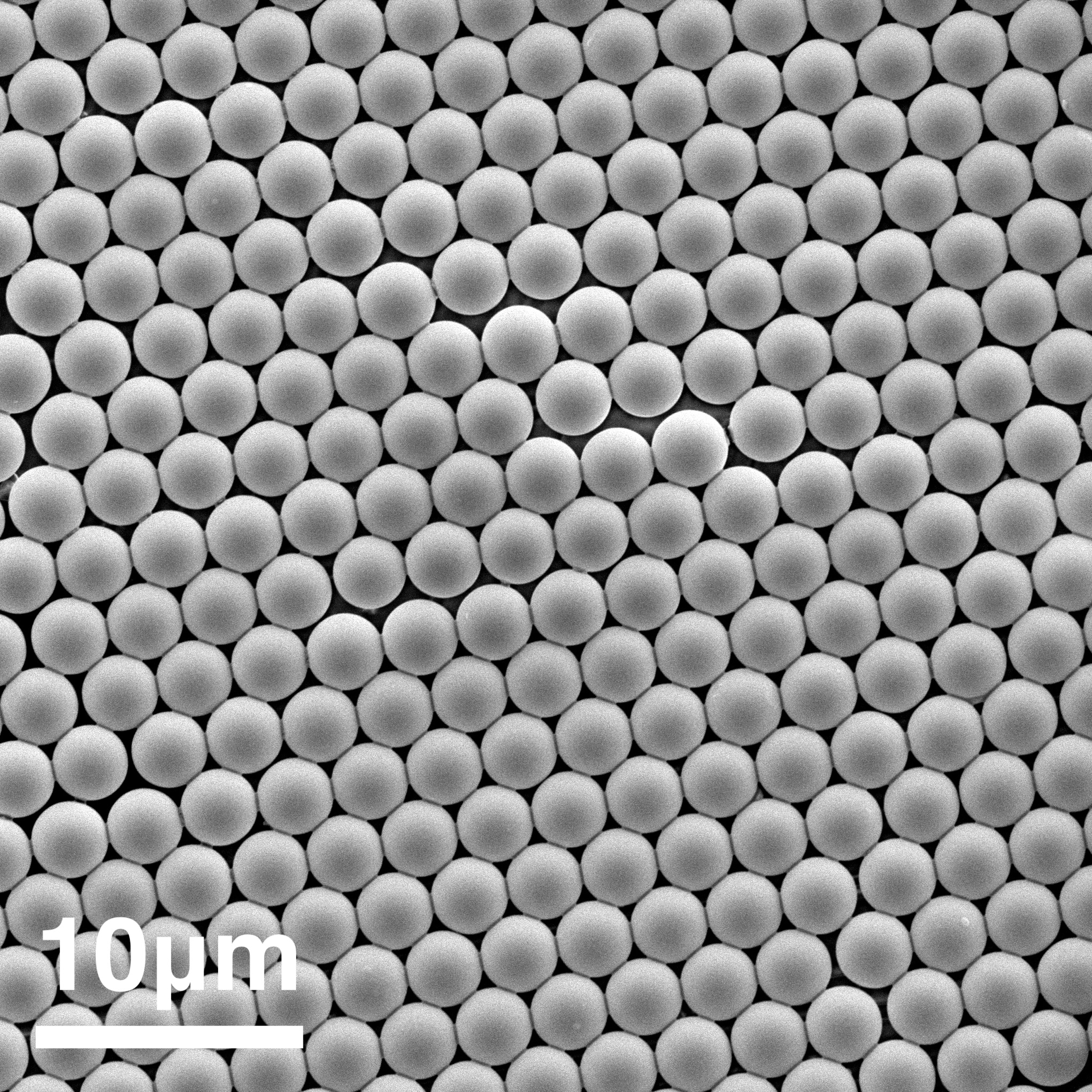}
        \caption{} \label{fig:sem_closed}
    \end{subfigure}
    \hfill
    \begin{subfigure}[b]{0.32\textwidth}
        \centering
        \includegraphics[width=\textwidth]{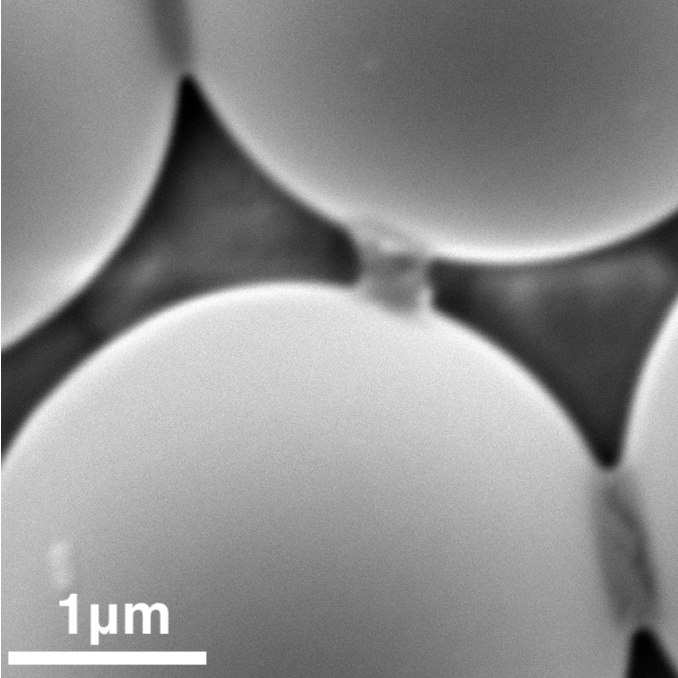}
        \caption{}\label{fig:hinges}
    \end{subfigure}
    \hfill
    \begin{subfigure}[b]{0.32\textwidth}
        \centering
        \includegraphics[width=\textwidth]{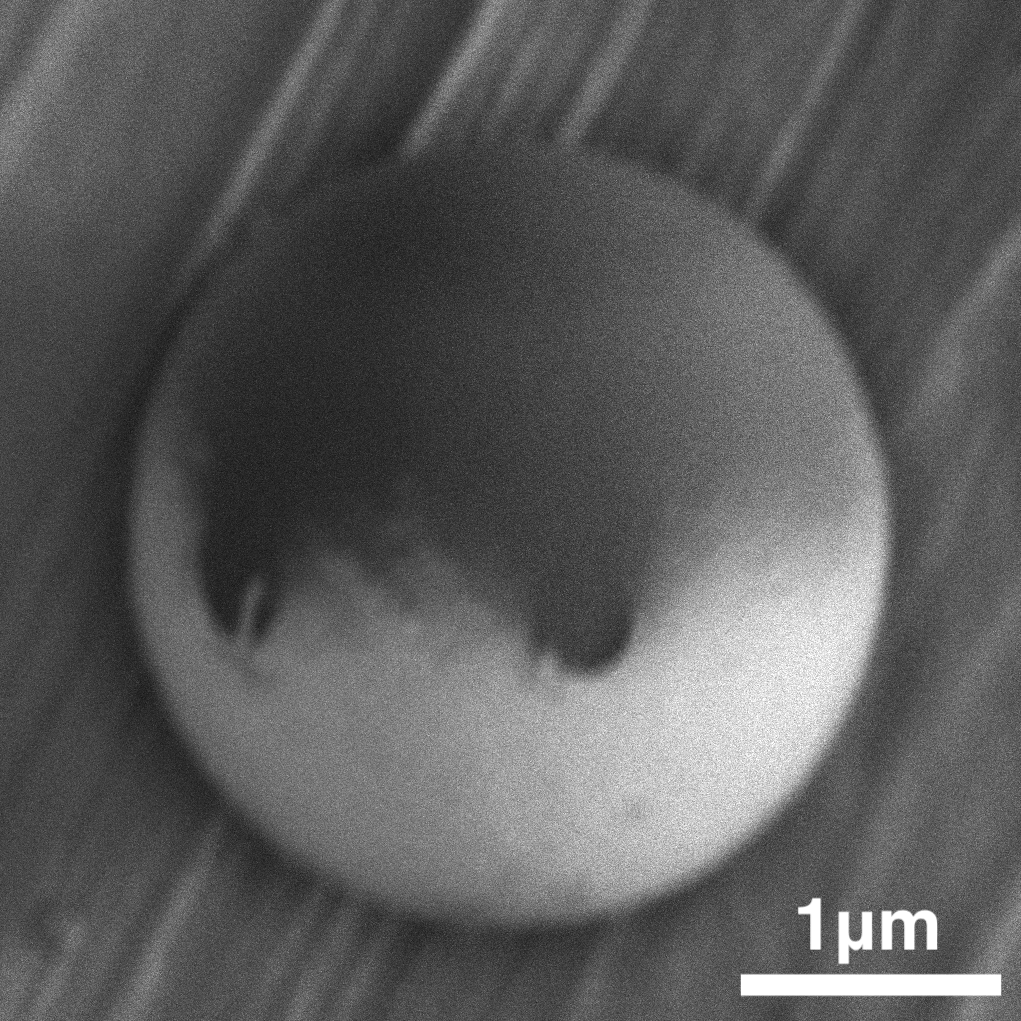}
        \caption{}\label{fig:closed_after}
    \end{subfigure}
    \caption{ 
        (a) Schematic of the self-electrophoresis hypothesis in \ce{Pt} Janus particles: the pole region acts as a cathode while the equator acts as an anode \cite{lyu2021active}. The waves around the equator zone of the particles indicate the shadow effect in the \ce{Pt} deposition due to neighboring particles. 
        (b) Schematic of a closed-packed monolayer on flat PDMS substrate termed as CM.
        (c) SEM Image of a closed-packed monolayer of \ce{SiO2} \qty{3}{\um} particles on CM substrate.
        (d) Zoomed in SEM image of particles shown in (c), the connection between the particles are contact point hinges resulted from sputtered metal layer on top of closed packed monolayer.
        (e) SEM image of a Janus microsphere detached from monolayer in (c). The curved region of deposition on the equator plane is due to the inter-particle contact point}
        \label{fig:schematic}
\end{figure}

Janus-particle fabrication begins with a close-packed colloidal monolayer (Figure~\ref{fig:sem_closed}) of silica (\ce{SiO2}) \qty{3}{\um} spheres on a flat PDMS stamp using the dry rubbing method~\cite{sotthewes2024toward} (details in Section~\ref{expsec}). Subsequently, the monolayer was then coated with a thin platinum layer (\qty{\sim 20}{\nm}) via sputtering to create Janus microspheres. This conventional Janus particle sample, prepared from a close-packed monolayer, is referred to as CM throughout the manuscript and serves as the control. Due to particle proximity and/or direct physical contact, we observed nanoscale bridges around the particle (Figure~\ref{fig:hinges}) and a curved equatorial deposit after release (Figure~\ref{fig:closed_after}). These bridges lead to a sharp increase in \ce{Pt} thickness at the contact points, resulting in a non-uniform thickness contrast (Figure~\ref{fig:fib_closed}) between pole and equator~\cite{zong_optically_2015}. Furthermore, the particles had a maximum of six bridges as they assembled into a hexagonal closed packing.

To obtain a proper, precise, and controllable Janus coating of inert particles by Physical Vapor Deposition (PVD), we introduced a novel fabrication method to create a particle monolayer with controlled particle spacing.

\begin{figure}[htbp]
    \centering
    \begin{subfigure}[c]{0.49\textwidth}
        \includegraphics[width=\textwidth]{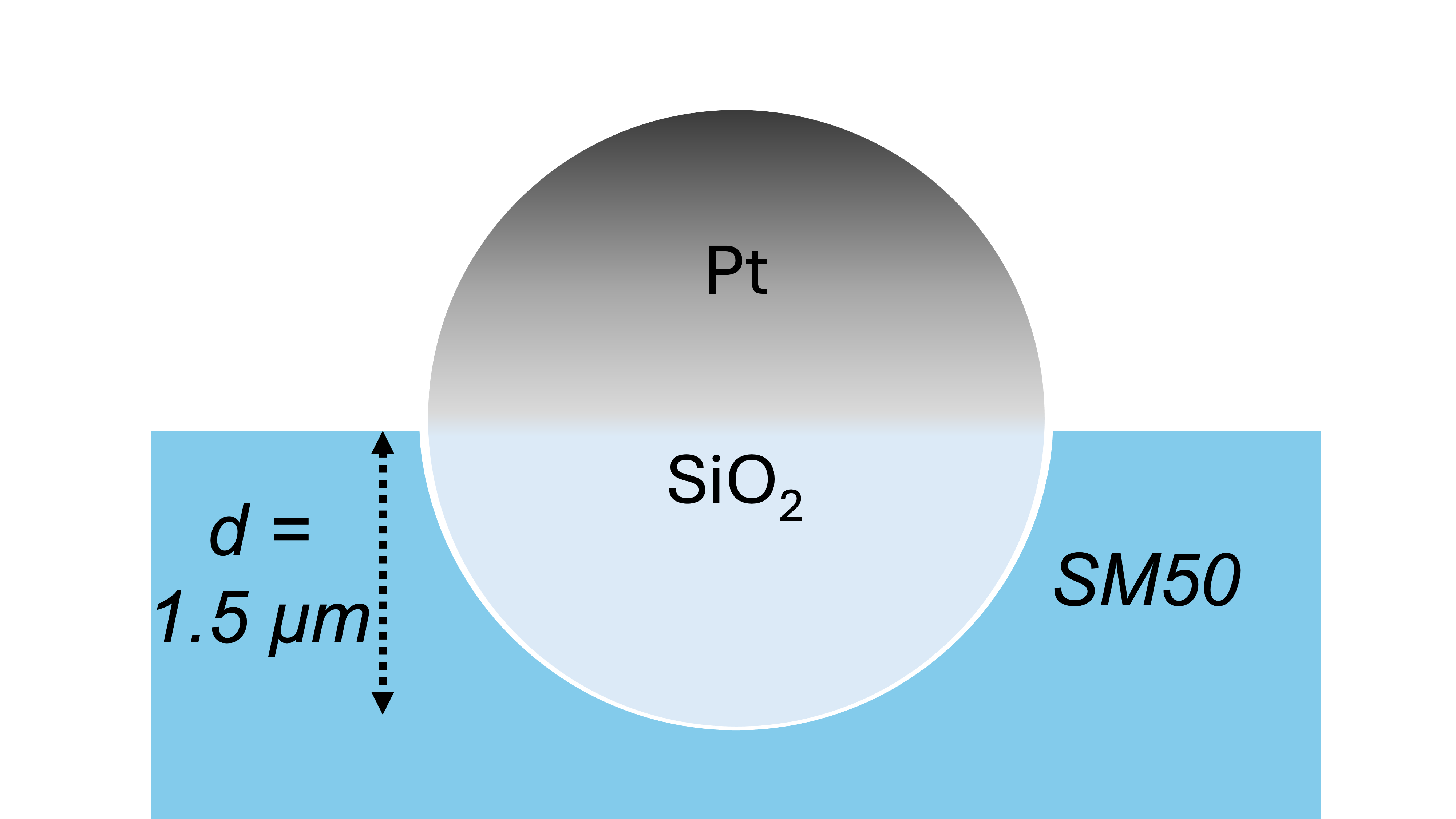}
        \caption{}\label{fig:schematic_SM50}
    \end{subfigure}
        \begin{subfigure}[c]{0.49\textwidth}
        \includegraphics[width=\textwidth]{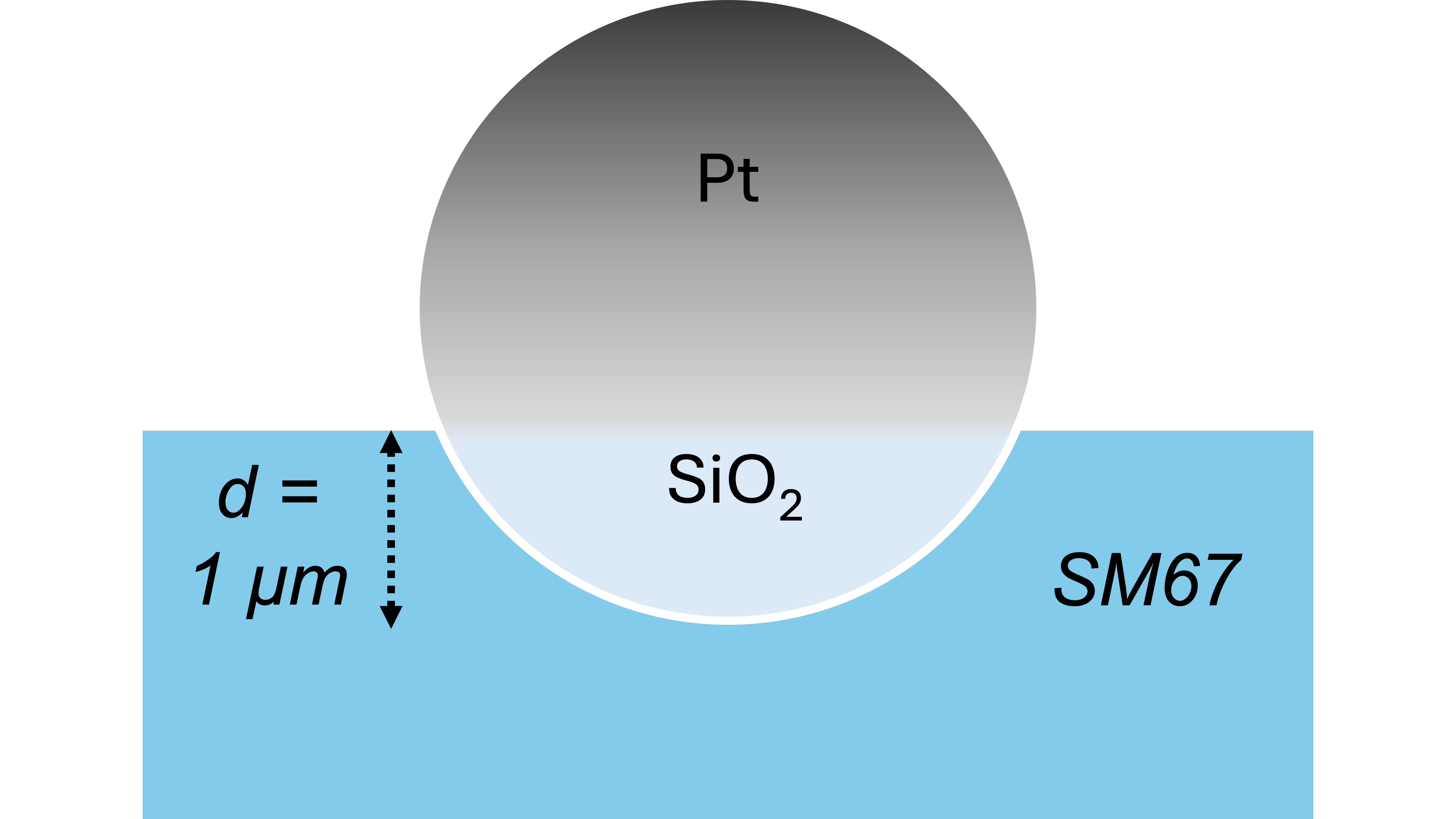}
        \caption{}\label{fig:schematic_SM67}
    \end{subfigure}

    \vspace{1.0em} 
    
    \begin{subfigure}[b]{0.32\textwidth}
        \includegraphics[width=\textwidth]{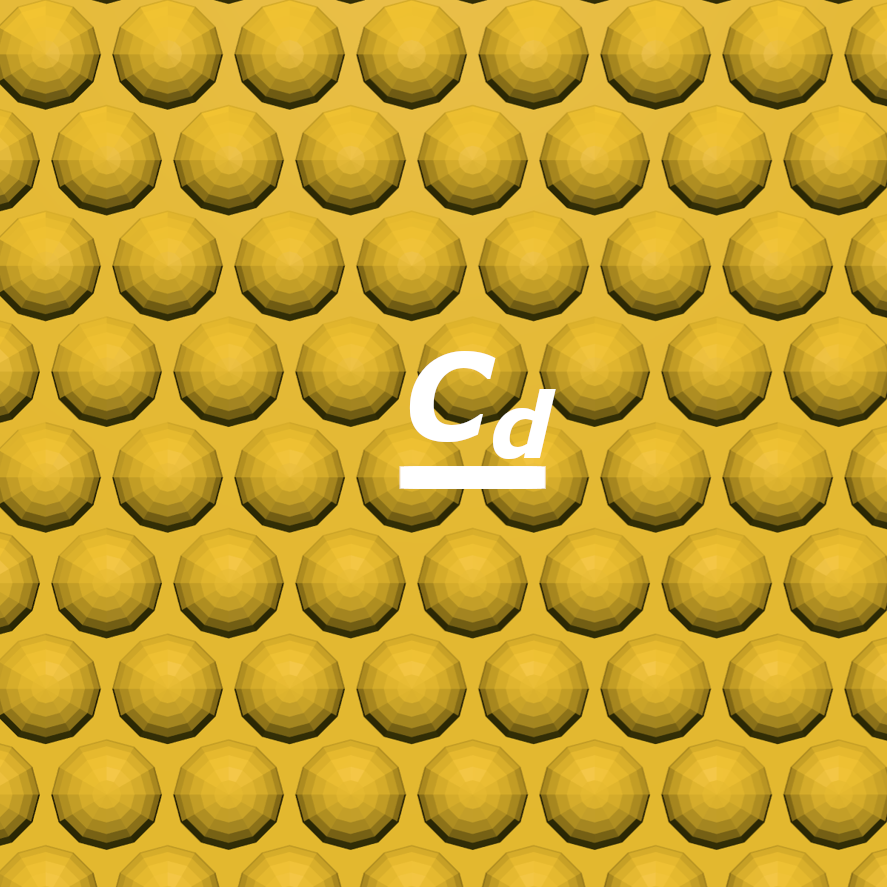}
        \caption{}\label{fig:grooves}
    \end{subfigure}
    \hfill
    \begin{subfigure}[b]{0.32\textwidth}
        \includegraphics[width=\textwidth]{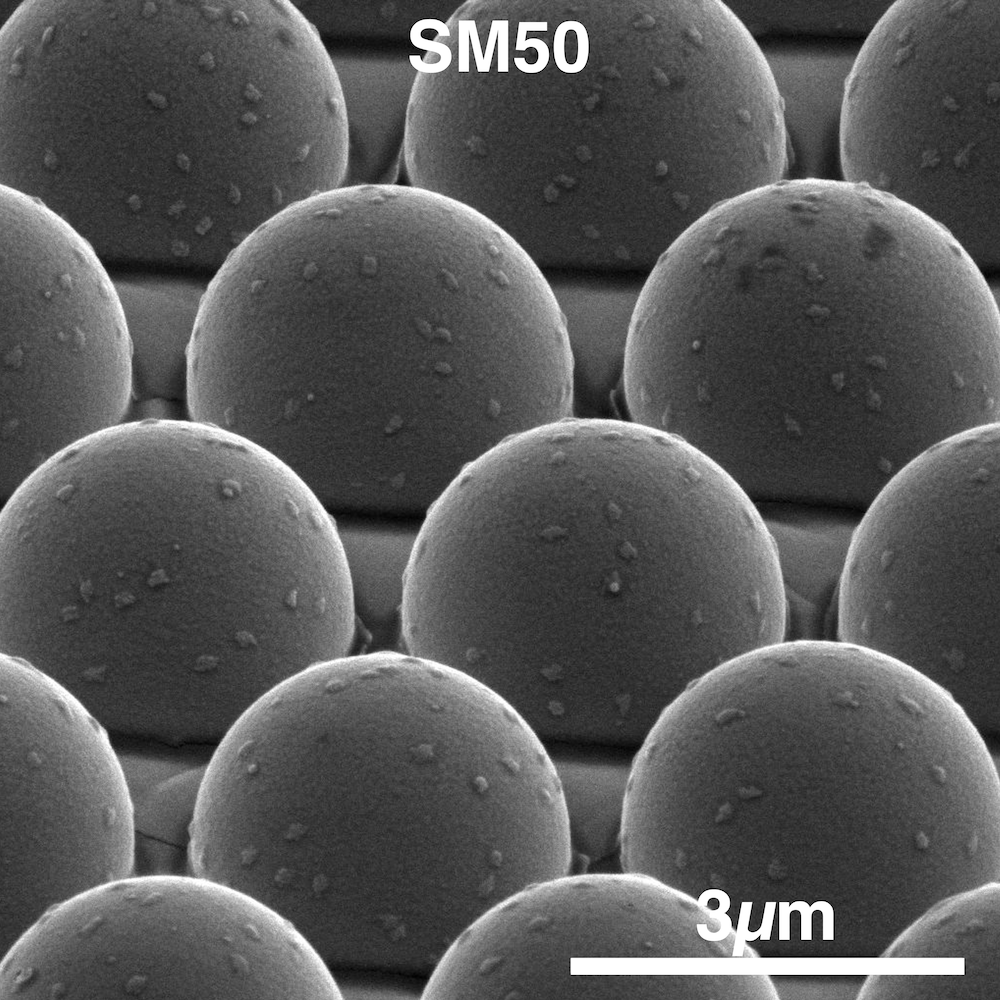}
        \caption{}\label{fig:sm50_angle}
    \end{subfigure}
    \hfill
    \begin{subfigure}[b]{0.32\textwidth}
        \includegraphics[width=\textwidth]{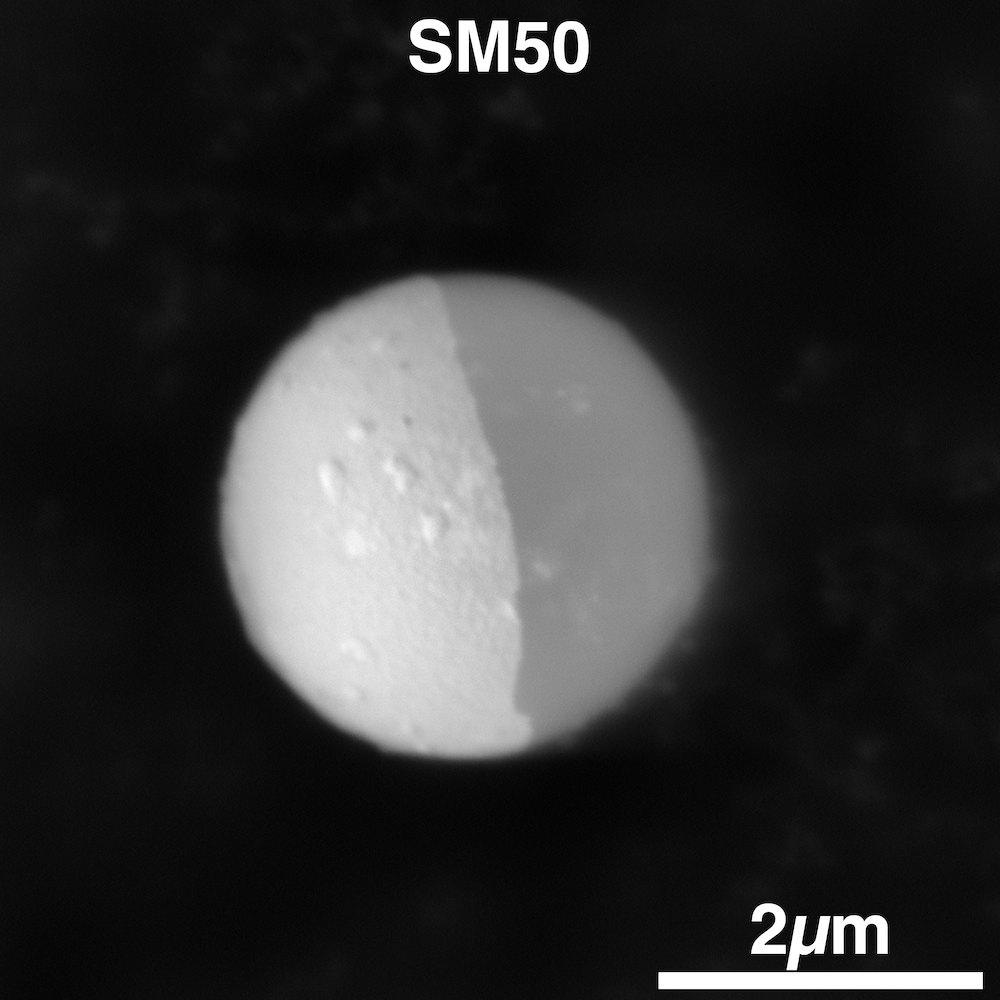}
        \caption{}\label{fig:sm50_after}
    \end{subfigure}

    \caption{ Schematic of microparticle embedded in a pattered substrate with groove depth:
    (a) \qty{1.5}{\um} equal to particle radius, named SM50; 
    (b) \qty{1}{\um}, named SM67; 
    (c) top view: design of spaced monolayer substrates with center-to-center spacing $c_d$; 
    (d) SEM image showing particles placement inside SM50 grooves, taken at \ang{52} tilt; 
    (e) SEM image of truly Janus particle prepared with SM50 substrate.}
    \label{fig:semcontrol}
\end{figure}

The method uses a specialized PDMS substrate with micro-patterns featuring circular grooves of depth $d$ and center-to-center spacing $c_d$, as shown in Figure~\ref{fig:grooves}. We carefully choose  $d\sim$~\qty{1.5}{\um} and center-to-center distance $c_d\sim$~\qty{3.5}{\um} such that a spaced monolayer of half-embedded particles was formed. $c_d$ was kept slightly greater than the particle diameter to avoid bridge formation, as well as to maximize the number of particles on the substrate. Groove radius \qty{1.6}{\um} was kept slightly larger than the particle radius \qty{1.5}{\um} to ease subsequent detachment of the particles from the substrate for motion study. This substrate was termed SM50, where SM stands for spaced-monolayer design, and 50 represents the percentage of the surface area covered by \ce{Pt}. Scanning electron microscopy (SEM) confirms that particles in the patterned grooves remain physically separated (Figure~\ref{fig:sm50_angle}). No metallic bridges were observed after \ce{Pt} sputtering and detachment of particles from the substrate (Figure~\ref{fig:sm50_angle}). The same spacing control was also achieved with larger center-to-center distances $c_d\sim$~\qtylist[list-units = single]{4.0;4.5}{\um} (see Supplementary material). These results show that our method is highly robust for assembling a monolayer of microparticles with a desired lattice spacing and eliminating contact-point defects. Throughout this manuscript, we will focus on results obtained using $c_d\sim$~\qty{3.5}{\um}.

\subsection{\textbf{Tuning particle embedding depth modulates \ce{Pt} thickness asymmetry}}

The self-electrophoresis hypothesis makes a concrete demand on the experiment: the pole-to-equator \ce{Pt} profile must be varied while the fuel concentration is held fixed. We therefore compare three embedding geometries that vary the extent to which each \qty{3}{\um} silica sphere is exposed to the sputter flux. Groove depth $d$ determines how deeply each sphere is embedded and therefore the fraction of its surface exposed to the incoming \ce{Pt} flux during sputtering (Figure~\ref{fig:semcontrol}). From the previous section, we have substrates CM and SM50, which have groove depths of \qtylist[list-units = single]{0.0; 1.5}{\um}, resulting in conventional, process-dependent hemispherical coatings and 50\% surface coating, respectively. We choose a groove depth of \qty{1.0}{\um} (Figure~\ref{fig:schematic_SM67}), which ideally leads to 66.67\% surface area covered (SM67) (refer to the supplementary material for the calculation of area).

\begin{figure}[htbp]
    \centering
    \begin{subfigure}[b]{0.32\textwidth}
        \includegraphics[width=\textwidth]{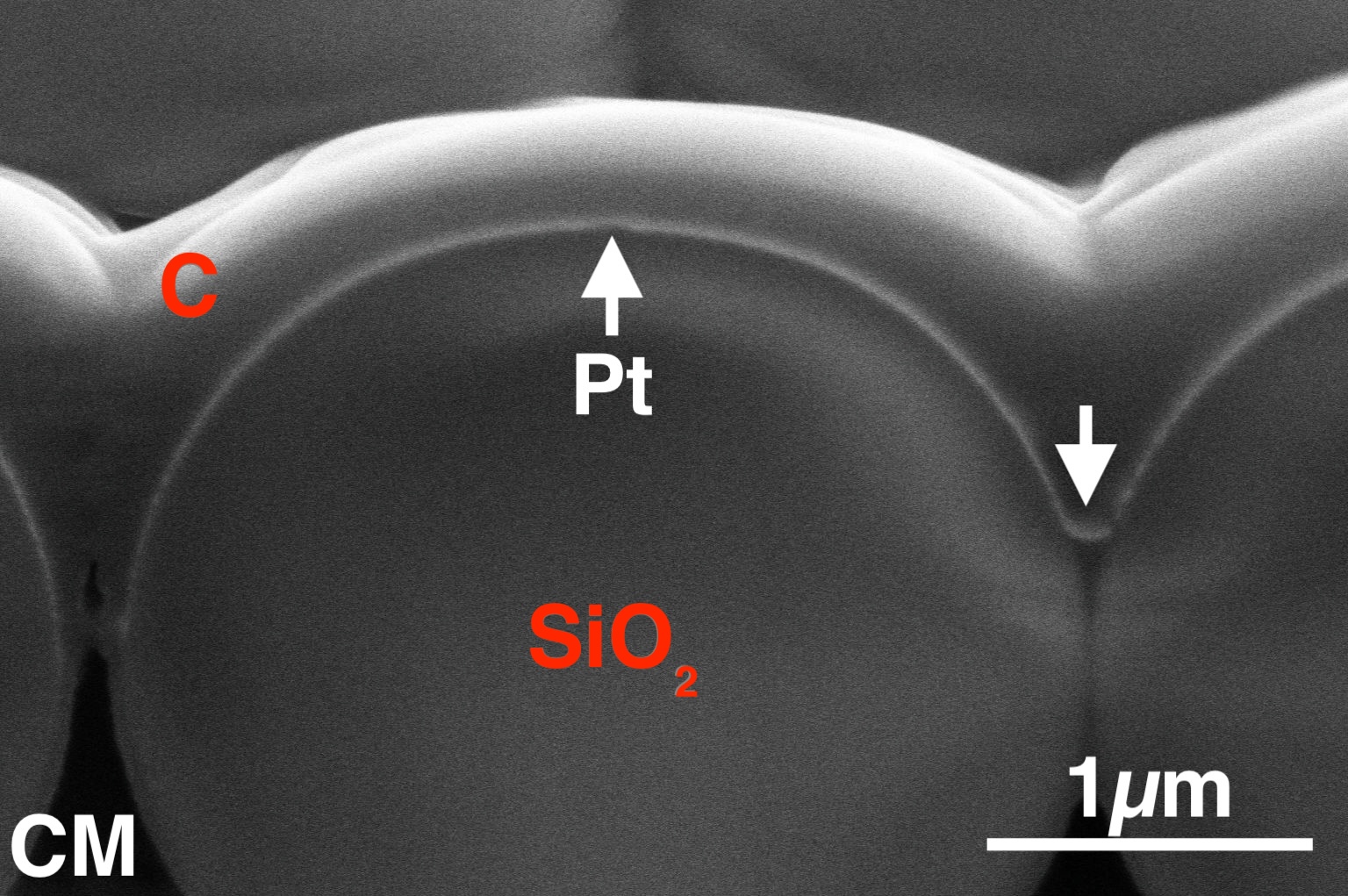}
        \caption{}\label{fig:fib_closed}
    \end{subfigure}
    \hfill
    \begin{subfigure}[b]{0.32\textwidth}
        \includegraphics[width=\textwidth]{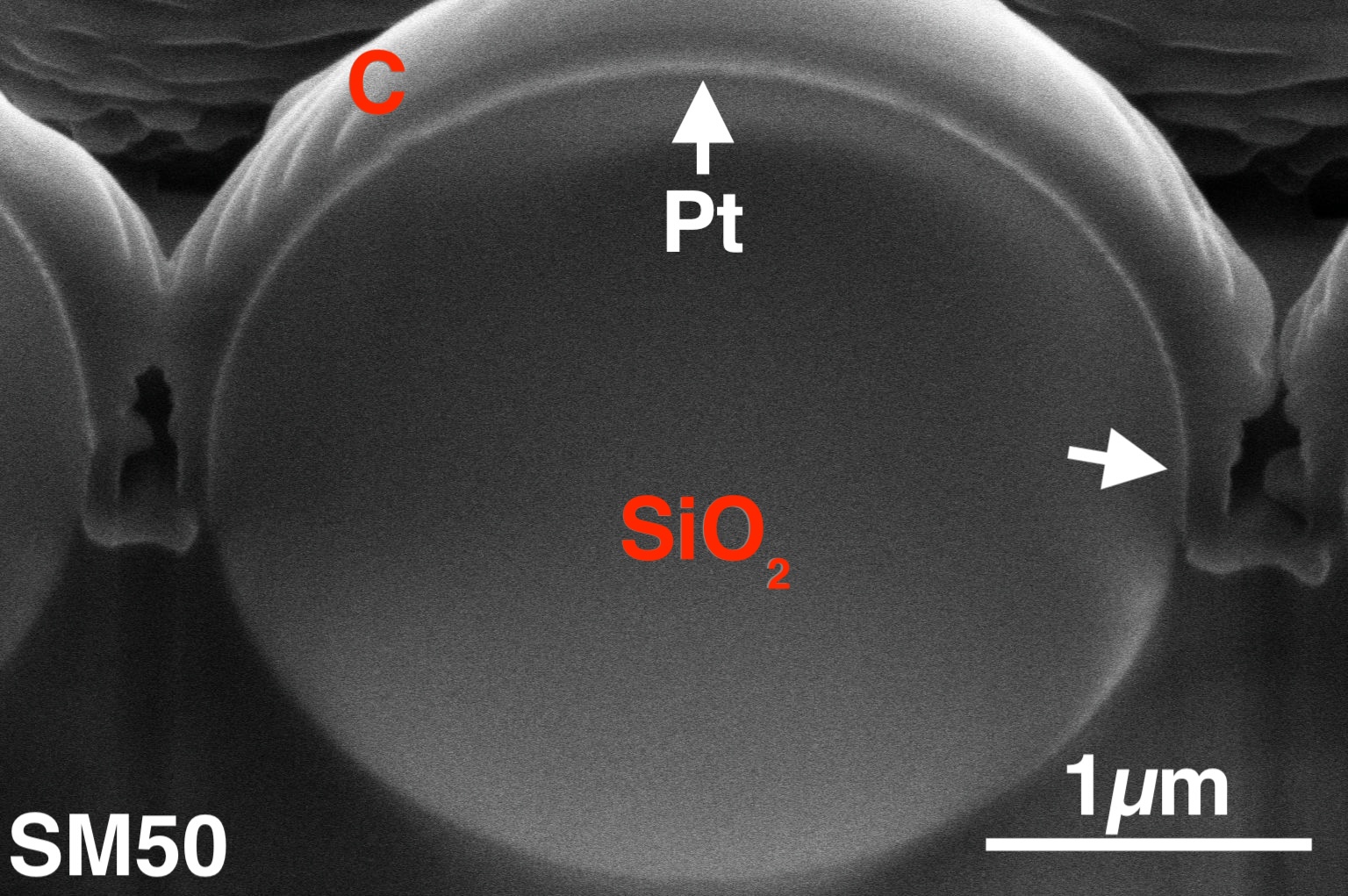}
        \caption{}\label{fig:fib_sm50}
    \end{subfigure}
    \hfill
    \begin{subfigure}[b]{0.32\textwidth}
        \includegraphics[width=\textwidth]{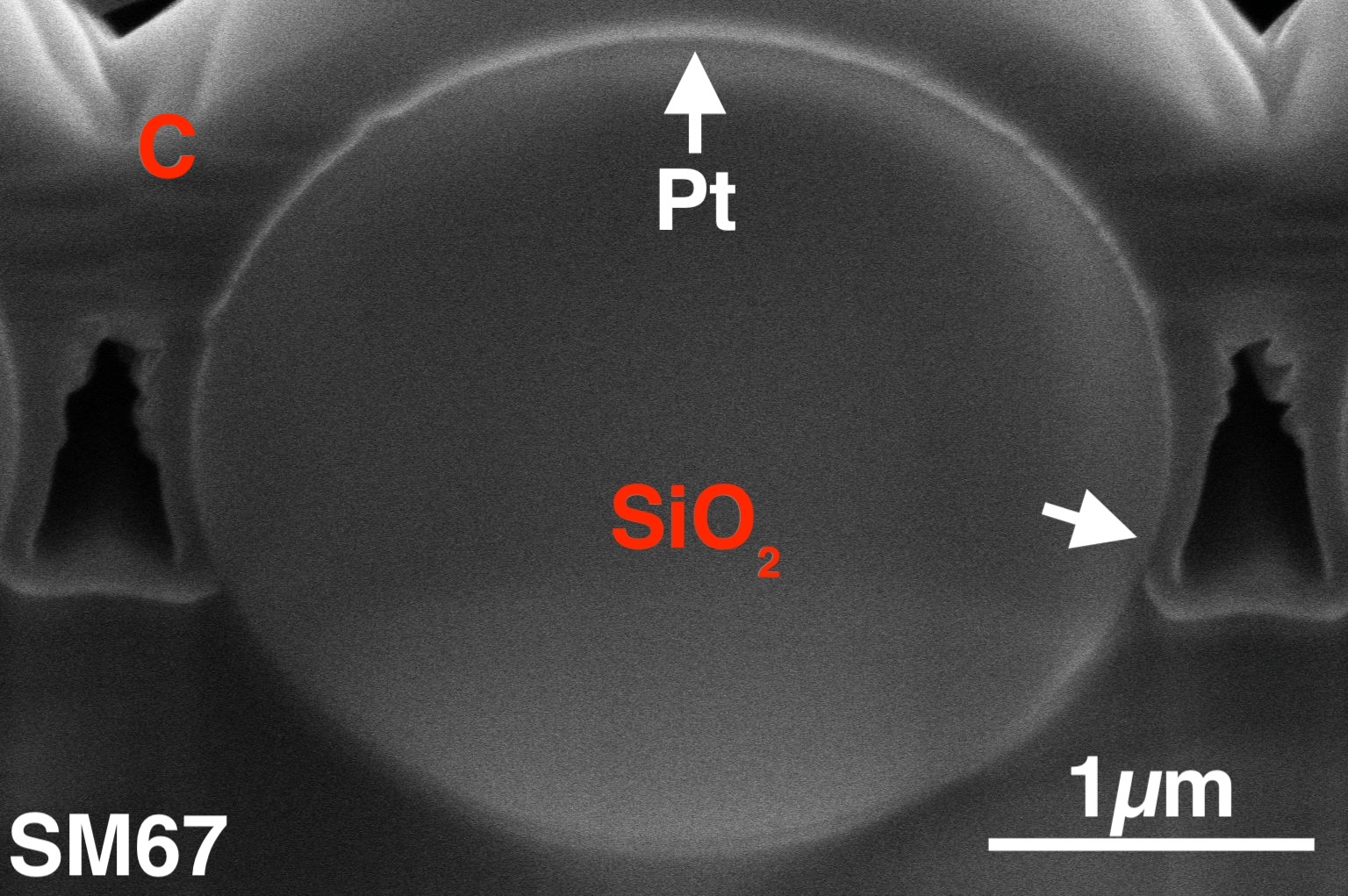}
        \caption{}\label{fig:fib_sm67}
    \end{subfigure}

    \caption{Particle cross-section obtained using Focused Ion Beam (FIB) milling for particles in: 
    (a) closed packing monolayers CM, 
    (b) SM50 substrate, and 
    (c) SM67 substrate. 
 In each image, C indicates a carbon protective layers deposited on the samples to avoid any \ce{Pt} damage during ion milling. 
    White arrows are for guidance of \ce{Pt} layer and \ce{SiO2} is silica microparticle .}
    \label{fig:fib}
\end{figure}

Next, Focused-ion-beam (FIB) cross-sections were prepared to examine the resulting \ce{Pt} thickness profiles (Figure~\ref{fig:fib}). In all three geometries, the coating is thicker at the pole than near the equator, consistent with the line-of-sight nature of sputtering~\cite{rashidi_local_2018}. The visual contrast between pole and equator appears more pronounced for particles prepared on the SM67 substrate than for those on the SM50 substrate. Therefore, \ce{Pt} thickness difference between pole and equator is greater in SM67 than in SM50. Because the absolute thickness difference approaches the SEM's resolution limit, the comparison remains qualitative. The two spaced geometries therefore differ primarily in embedding depth, and with it in the intended \ce{Pt} thickness profile. They also differ in the coated-area fraction; that second variable is addressed in the Discussion.

\subsection{\textbf{Propulsion speed increases with designed thickness asymmetry}}
\begin{figure}[htbp]
    \centering
    \begin{subfigure}[b]{0.49\textwidth}
        \includegraphics[width=\textwidth]{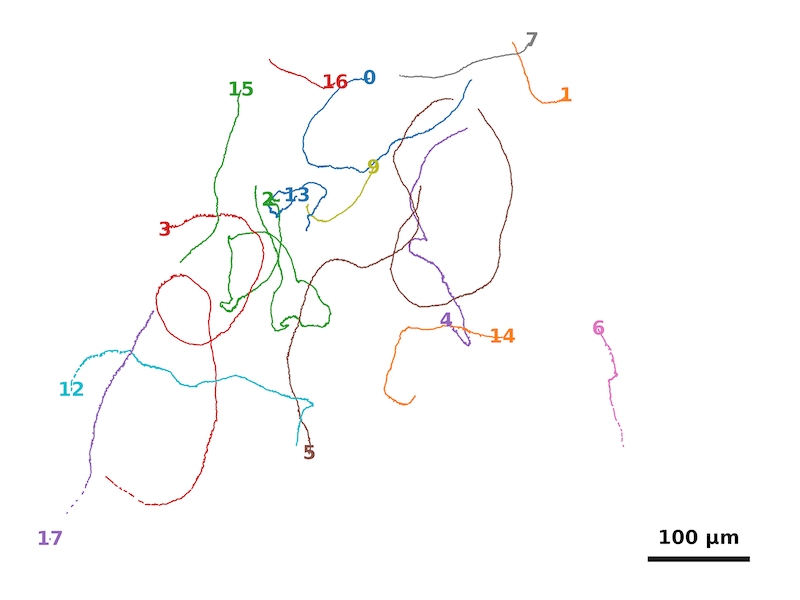}
        \caption{}
    \end{subfigure}
    \hfill
    \begin{subfigure}[b]{0.49\textwidth}
        \includegraphics[width=\textwidth]{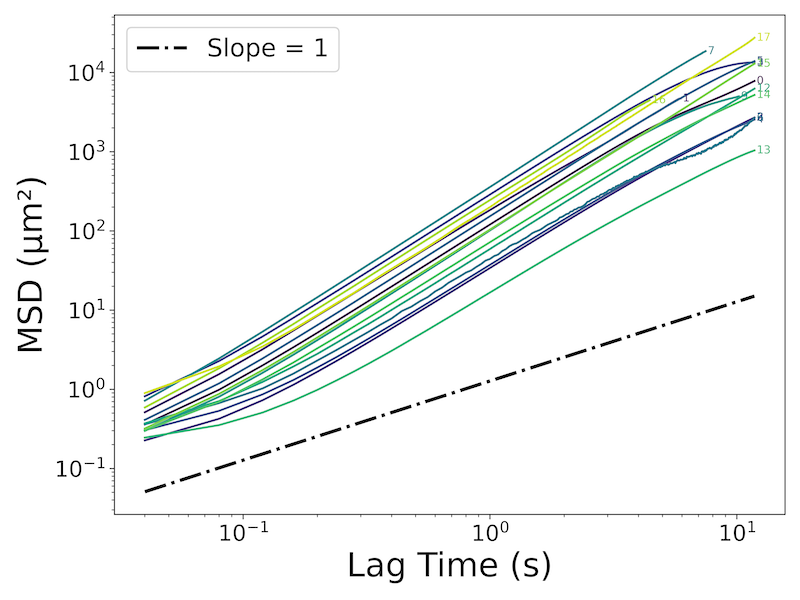}
        \caption{}
    \end{subfigure}

    \caption{Particle tracking analysis: 
    (a) Trajectories for SM67 substrate at 2.5\% hydrogen peroxide concentration for 120 seconds. 
    (b) Logarithmic scale mean squared displacement (MSD) of these tracks, black dotted line is slope 1 line drawn for reference.}
    \label{fig:particle_tracking_combined}
\end{figure}

The Janus particles from the CM, SM50, and SM67 substrates were detached in ultrapure water using ultrasonic pulses, and their activity was assessed in 2.5\% \ce{H2O2}. The positions of the particles in the $xy$ plane were tracked using the \textit{trackpy} library in Python\cite{allan_soft-mattertrackpy_2025}, and their mean squared displacements (MSD) and velocities were calculated using custom-built codes in Python \cite{brar_microparticle_2026}. Figure~\ref{fig:particle_tracking_combined} depicts the analyzed trajectories and MSDs of the particles detached from the SM67 substrate. We divide the particles into three categories: stuck, intermittent, and continuously moving (see Section~\ref{es:tracking}). The fraction of stuck or intermittently moving particles decreased from 4.8\% (CM) to 3.7\% (SM50) and 2.4\% (SM67), consistent with the elimination of bridges and discontinuities at the particle equator.

\begin{figure}[htbp]
    \centering
    \begin{subfigure}[b]{0.49\textwidth}
        \includegraphics[width=\textwidth]{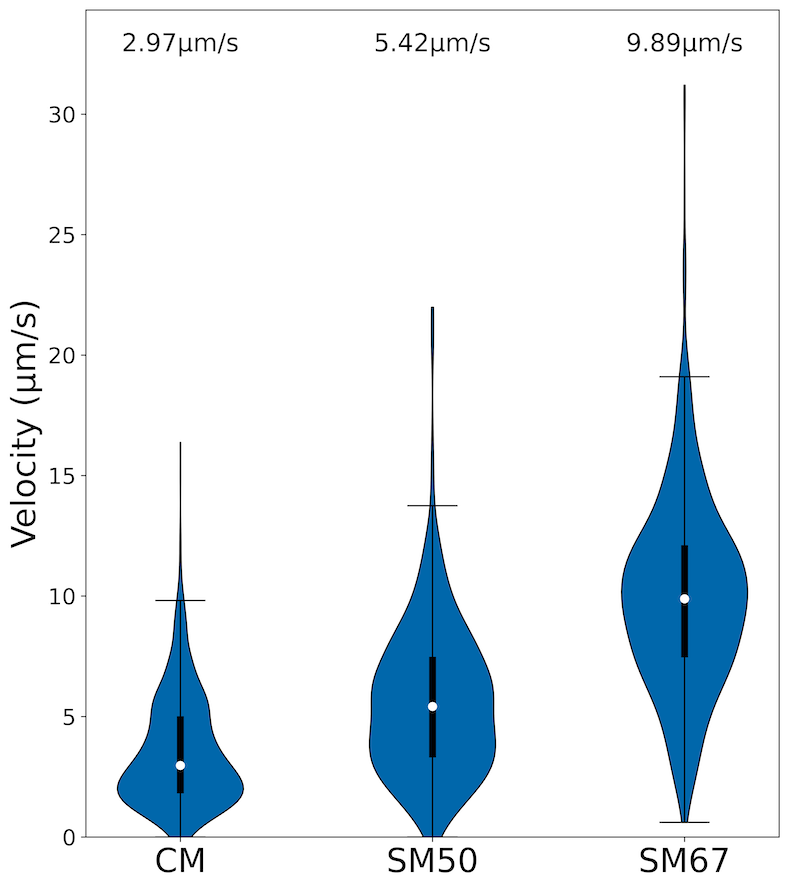}
    \end{subfigure}
    \caption{Distributions of self-propulsion velocities of particles obtained with CM, SM50, and SM67 mask. The white dot represents the median velocity of the distributions, with the exact values reported at the top of each distribution, showing an increasing trend. $p$-test values for the velocity comparison between three cases are \num{\ll 0.0001}, indicating the statistical significance of the observed differences.}
    \label{fig:3violin}
\end{figure}

Next, we choose only continuously moving particles for velocity analysis. Each particle's propulsion velocity was extracted from fitting the equation (see section~\ref{es:tracking}) to its MSD data. Figure~\ref{fig:3violin} reports the velocity distribution for more than 500 particles in each condition. There is a clear trend in median propulsion speed $v_{\text{CM}} < v_{\text{SM50}} < v_{\text{SM67}}$. In CM samples, the speed of the active particles was comparable to that reported in literature~\cite{choudhury_surface_2015} but lower than that of spaced monolayer particles. Furthermore, in spaced particles, the SM67 particles were strikingly faster than the SM50 particles. The counterintuitive observation that a non-50-50 Janus geometry leads to higher self-propulsion velocity is indeed fully consistent with self-electrophoretic propulsion: the larger metal-thickness difference in SM67 generates a stronger local electric field and therefore faster directed motion~\cite{lyu2021active}. The difference between each pair is statistically significant as pointed by a $p < 10^{-4}$ in the Mann-Whitney $U$ test \cite{mann_test_1947, wilcoxon_individual_1945}. Furthermore, velocities are differently distributed in all cases. In the CM sample, the distribution is skewed towards lower velocities, whereas in SM67 it is skewed towards higher velocities and shifts towards a normal distribution.

\section{Discussion}\label{discussion}
The central experimental observation of this work is a systematic increase in median propulsion speed equation for \ce{Pt-SiO2} Janus particles prepared on closed-monolayer (CM), spaced-monolayer with 50\% \ce{Pt} coverage (SM50), and spaced-monolayer with roughly 67\% \ce{Pt} coverage (SM67) substrates. Particles fabricated on CM substrates are the slowest and exhibit the highest fraction of stationary or intermittently moving particles. This behavior is consistent with the presence of metallic nano-bridges, which can act as physical hinges, thereby introducing frictional interactions with the container surface. Elimination of these bridges in the spaced monolayers (SM50 and SM67) already increases the yield of continuously moving particles.

In the absence of contact-point bridges, the additional speed increase from SM50 to SM67 correlates with a visually more pronounced \ce{Pt} thickness difference, as revealed by FIB cross-sections of particles prepared on these substrates. These results align with models in which a nanoscale thickness contrast across the \ce{Pt} cap generates self-sustaining ionic currents and the associated electric field that drives self-electrophoresis~\cite{brownIonicEffectsSelfpropelled2014a, ebbens_electrokinetic_2014, lyu2021active}. In the simplest picture, the thicker polar region and thinner equatorial/bottom region function as differentiated electrodes; increasing the thickness difference is expected to amplify the effective electric field. The present fabrication platform enables tuning of this difference through controlled physical embedding of particles in the patterned SM50 and SM67 substrates. Although the thickness profiles are visibly distinct in FIB–SEM images, the analysis remains qualitative: the absolute thickness difference is near the instrument's resolution limit. Moreover, the method inherently alters the fraction of the particle surface exposed to the \ce{Pt} coating. Consequently, the observed speed trend cannot yet be attributed solely to thickness asymmetry.

Nevertheless, several observations indicate that the thickness difference plays a major role in the velocity trend. First, SEM images of CM particles (Figure~\ref{fig:closed_after}) show that the lack of efficient shadowing effect of neighboring particles increases \ce{Pt} coverage below the equator, resulting in a slightly higher overall coverage on CM than on SM50. If pure coverage, rather than thickness difference, were the dominant factor controlling propulsion, a higher velocity would be expected for CM relative to SM50; the opposite of the observed trend.  Literature further argues against a simple coverage-area-driven speed increase beyond $~50\%$, rendering a pure area effect unlikely to explain the SM50 to SM67 increase~\cite{liuQuantifyingUnderstandingTilt2025, haroonCoupledInterfacialPhenomena2026, mengEffectParticleSize2022}. Second, the reaction rates on nanometre-scale Pt films are known to vary strongly with thickness in the \qtyrange[list-units = single]{1}{10}{\nm} regime, and the natural thickness gradient produced by directional deposition has been identified as a plausible origin of the anodic/cathodic differentiation required for self-electrophoresis~\cite{brownIonicEffectsSelfpropelled2014a, ebbens_electrokinetic_2014, archerPickeringEmulsionRoute2018}. Third, near the inter-particle contact points of CM particles, this \ce{Pt} difference between pole and equator is reduced relative to the remainder of the surface, producing a non-uniform thickness difference (Figures~\ref{fig:closed_after} and \ref{fig:fib_closed}) that compounds the frictional effects of the metallic bridges. Hence, with inter-particle bridges eliminated and coverage-area effects expected to be negligible beyond $~50\%$, the higher velocity of SM67 compared to SM50 is most readily explained by the enhanced pole–equator thickness difference.

Beyond fundamental insight, the patterned-substrate method offers practical advantages over conventional close-packed monolayers. Inter-particle spacing can be tuned, metallic bridges are eliminated, and the degree of particle embedding, and therefore the coated-area fraction, can be set by design. These features should prove useful for both the fabrication of truly Janus particles and for applications such as active-particle-based microrobots for drug delivery, where low fuel concentration and rapid motion are desirable~\cite{sanchez_chemically_2015}. Alternative routes to faster Janus particles, such as non-spherical shapes \cite{maric_shape-controlled_2024}, which often compromise scalability or introduce strong rotational components not desirable for microparticle-based microrobots. The present approach retains spherical symmetry while delivering a measurable speed gain and improved scalability.

\section{Conclusions}\label{concl}
By replacing the conventional close-packed particle monolayer with geometrically defined spaced monolayers, we achieve simultaneous control over inter-particle separation and \ce{Pt} thickness asymmetry. 
The resulting active Janus microparticles display a clear hierarchy of propulsion speeds ($v_{\rm CM} < v_{\rm SM50} < v_{\rm SM67}$) at fixed fuel concentration. 
This trend is consistent with expectations from \ce{Pt} thickness-gradient-driven self-electrophoresis and thereby supplies further indirect evidence that self-electrophoresis is the operative propulsion mechanism of such Janus colloids. 
The patterned-substrate approach additionally eliminates metallic nano-bridges, retains axial symmetry, and remains readily scalable, offering a practical route to both fundamental studies of microscale active motion and to applications such as fuel-efficient microrobots for targeted delivery.

\section{Experimental Section}\label{expsec}
\subsection{Materials}
Commerically available \ce{SiO2} microspheres (diameter \qty{3.0(0.1)}{\um}) from Sigma-Aldrich were used as received. Sylgard 184 silicone elastomer kit (Dow Corning), trichloro(1H, 1H, 2H, 2H-perfluorooctyl)silane (97\%, Sigma-Aldrich), IP-S photoresist (Nanoscribe), propylene glycol monomethyl ether acetate (PGMEA, Sigma-Aldrich), and hydrogen peroxide (30 wt\%, Sigma-Aldrich) were used without further purification. \qty{18.2}{M\Omega\cdot cm} Ultrapure water (Milli-Q EQ 7008/7016) was used throughout.

\subsection{Design and Fabrication of the Master}\label{es:design}
Master structures consisting of hexagonal arrays of hemispherical domes were designed in Blender and converted to STL files. Two geometries were prepared, with center-to-center spacing of \qty{3.5}{\um}. The dome heights were set to \qty{1.0}{\um} and \qty{1.5}{\um}, producing PDMS grooves of corresponding depths after replication. The dome base radius was chosen slightly larger than the particle radius (\qty{1.6}{\um}) to facilitate later particle release.

The masters were fabricated on SuperFrost Gold glass slides (Epredia) by two-photon polymerization (QuantumX, Nanoscribe GmbH). Substrates were cleaned by sequential ultrasonication in acetone, isopropanol, and water, followed by oxygen-plasma treatment (\qty{20}{W}, \qty{60}{s}, \qty{1.6e{-2}}{mbar}). IP-S photoresist was drop-cast and polymerized with a $25\times$ objective (NA 0.8) in dip-in mode using the following parameters: scan speed \qty{200}{mm/s}, laser power \qty{70}{mW}, slicing distance \qty{1.2}{\um}, hatching distance \qty{0.25}{\um}. After printing, samples were developed in PGMEA for \qty{20}{min} and rinsed in isopropanol for \qty{10}{min}.

\subsection{Preparation of Patterned PDMS Stamps}\label{es:pdms_stamp}
The masters were silanized with trichloro(1H, 1H, 2H, 2H-perfluorooctyl)silane vapor for \qty{72}{hrs}. Sylgard 184 base and curing agent were mixed in a 10:1 weight ratio, degassed, poured onto the master, degassed again, and cured at \qty{70}{\celsius} for \qty{4}{hrs}. The cured PDMS was carefully peeled to yield stamps containing arrays of spherical grooves (mentioned in section~\ref{es:design}). Flat PDMS stamps (no grooves) served as the Control substrate.

\subsection{Assembly of Colloidal Monolayers}\label{es:monolayer_assembly}
A \qty{50}{mg/mL} aqueous suspension of \ce{SiO2} particles was prepared and homogenized by vortexing and brief ultrasonication. \qty{25}{\uL} of the suspension was drop-cast onto a clean glass slide and allowed to dry completely at room temperature. The PDMS stamp (grooved side down) was then placed on the dried particle layer and gently rubbed in a circular motion under controlled pressure until a uniform monolayer was transferred onto the stamp. Multilayer regions (visible as darker areas) were eliminated by additional light pressure. The resulting monolayers were inspected by optical microscopy before metal deposition.
\begin{figure}[htbp]
    \centering
    \begin{subfigure}[b]{0.8\textwidth}
        \includegraphics[width=\textwidth]{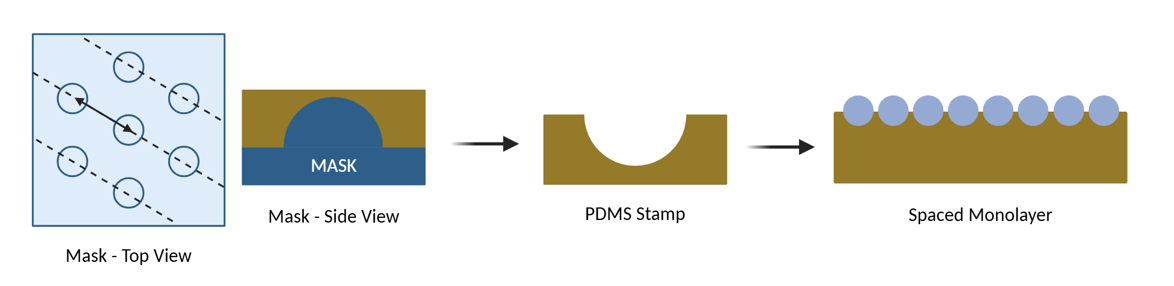}
    \end{subfigure}

    \caption{Schematic representation of Mask design with arrows pointing in the rubbing direction and Dry rubbing method. }\label{fig:schematic_methods}
\end{figure}
\subsection{Platinum Deposition and Particle Release}\label{es:janus}
Platinum was deposited by RF/DC magnetron sputtering (Kenosistec Srl) at Power \qty{70}{W}, deposition time \qty{20}{s}, working pressure of \qty{2e{-2}}{mbar} in \ce{Ar}, target–substrate distance \qty{7.5}{cm}. After deposition, the stamps were immersed in ultrapure water and subjected to mild ultrasonication to release the Janus particles. The resulting suspensions were used on the same day for motility experiments.

\subsection{Structural Characterization}\label{es:charaterization}
Scanning electron microscopy (SEM) was performed on the Benchtop SEM Phenom XL to assess monolayer quality and particle morphology. To characterize the pole-to-equator platinum thickness profile, selected particles were cross-sectioned by focused-ion-beam (FIB) milling on a dual-beam FIB-SEM system (Helios Nano Lab 600i (Thermo Fisher Scientific, Waltham, MA, USA)). Before the FIB cutting, a carbon protective pad was deposited over the region of interest in order to avoid any damage of the Pt layer. FIB cross sections were obtained with the \ce{Ga}\(^+\) beam working at \qty{30}{kV} and \qty[parse-numbers=false]{2.5 - 0.43}{nA} and impinging perpendicularly on the sample surface. SEM images of the cross sections were acquired with an Everhart-Thornley Detector (ETD) working at \qty{5}{kV} accelerating voltage and \qty{43}{pA} landing current, using secondary electrons detection, while keeping the sample tilted at \ang{52} and \qty{\sim4.1}{mm} working distance.

\subsection{Motility Assays}\label{es:motility}
All experiments were performed at room temperature. \qty{10}{mL} of \qty{2.5}{vol\%} \ce{H2O2} was placed in a plastic petri dish of diameter \qty{5.2}{cm} and allowed to equilibrate for a few minutes to minimize convective drifts. \qty{200}{\uL} of the particle suspension was then added, and particles were allowed to sediment. Motility was recorded on a digital optical microscope (Hirox HRX-01 3D) at $600\times$ magnification and $25$ frames per sec for \qty{120}{s}. For each condition, five videos were acquired at different locations. At least three independent fabrication batches were examined for every substrate type.

\subsection{Particle Tracking and Data Analysis}\label{es:tracking}
Videos were analyzed using the trackpy library \cite{allan_soft-mattertrackpy_2025} and custom Python scripts [ref]. Particle centroids were detected in each frame and linked into trajectories. Trajectories shorter than $10$\si{s} were discarded. Particles were classified as “stuck” if their displacement remained below one particle radius for the entire recording, and as “non-continuous” if they exhibited intermittent immobilization lasting \qty{\geq5}{s}. Only continuously motile particles were retained for velocity analysis.

Residual drift was subtracted by calculating the mean positions of all particles. Mean-squared displacements (MSDs) were computed and fitted at short lag times \cite{bailey_fitting_2022} to
\[
\mathrm{MSD}(\tau) = 4D_T\tau + \frac{2v^2}{D_R^2}\bigl(D_R\tau - 1 + e^{-D_R\tau}\bigr)
\]
where $D_T$ was fixed to the value measured for passive (non-catalytic) particles of the same size in water, and $D_R$ was obtained from the Einstein relation $D_R = 3D_T/R^2$. Propulsion speed $v$ was extracted as a free parameter. Statistical significance between conditions was assessed by the Mann-Whitney $U$ test \cite{mann_test_1947, wilcoxon_individual_1945}.

\section{Acknowledgment}
The authors thank Aliria Poliziani and Dr. Hilda Gomez Bernal for their technical help during the experiments and instrument training.

\section{Funding Statement}
This project has received funding from the European Research Council (ERC) under the [European Union’s Horizon 2020 research and innovation programme][European Union’s Horizon Europe research and innovation programme] (Grant agreement No. [948590]).

\newpage
\appendix
\section{Increasing interparticle distance in the spaced substrate}\label{appendix:cd}
\begin{figure}[H]
    \centering
    \begin{subfigure}[b]{\columnwidth}
        \centering
        \includegraphics[width=0.32\textwidth]{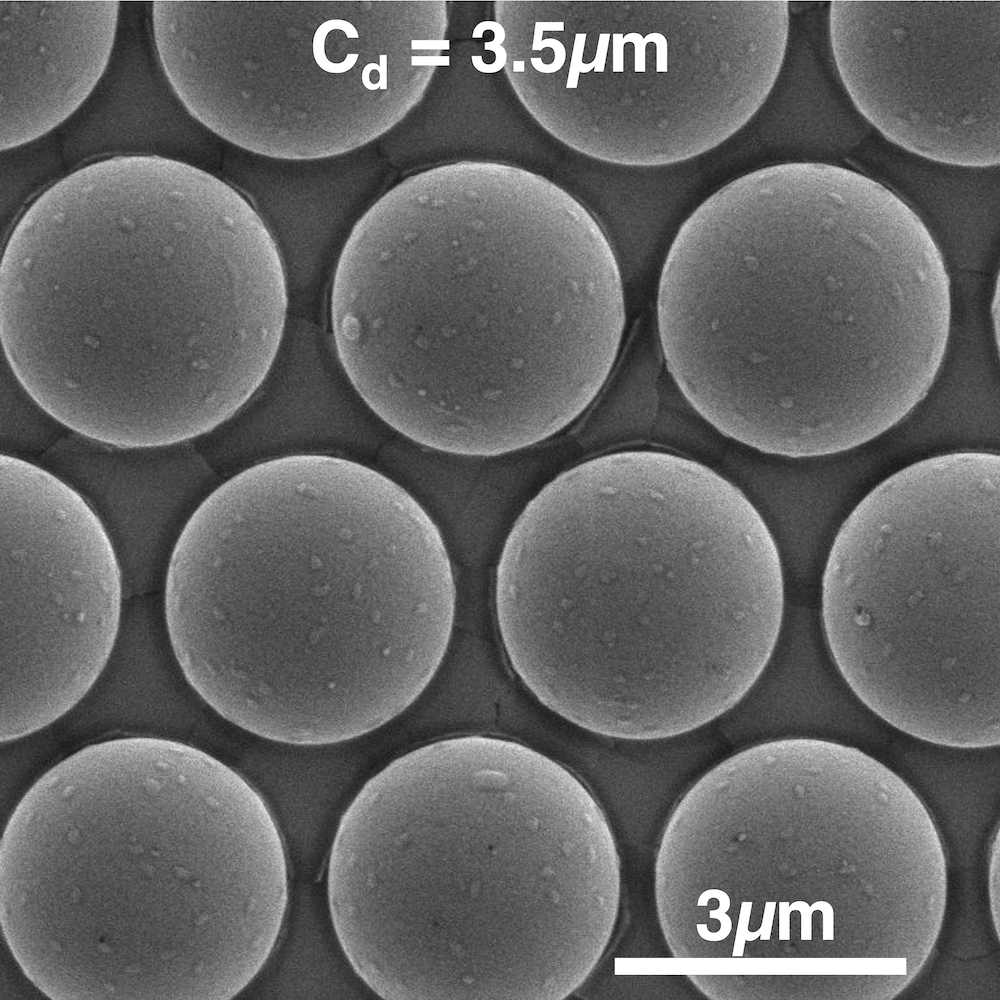}
        \caption{}\label{fig:cd35}
    \end{subfigure}
    \newline
    \vspace{0.0em}
    \begin{subfigure}[b]{\columnwidth}
        \centering
        \includegraphics[width=0.32\textwidth]{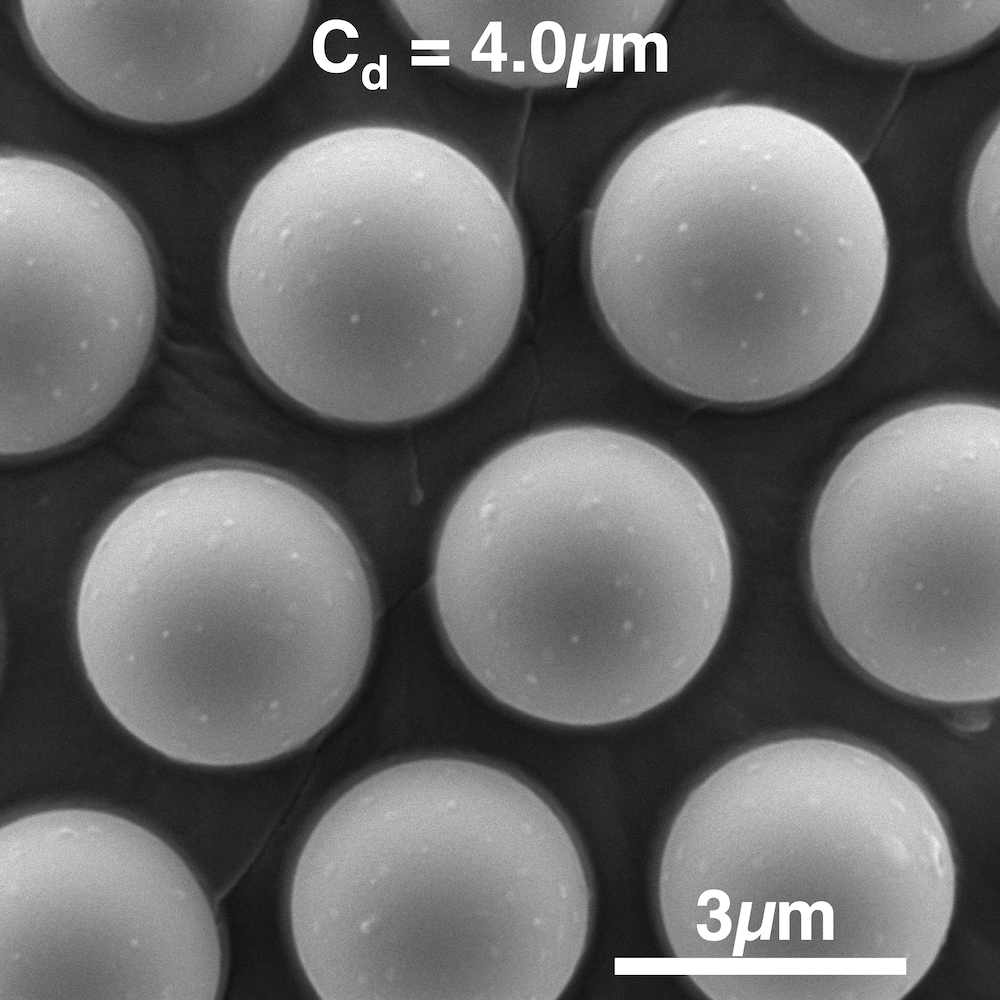}
        \caption{}\label{fig:cd40}
    \end{subfigure}
    \newline
    \vspace{0.0em}
    \begin{subfigure}[b]{\columnwidth}
        \centering
        \includegraphics[width=0.32\textwidth]{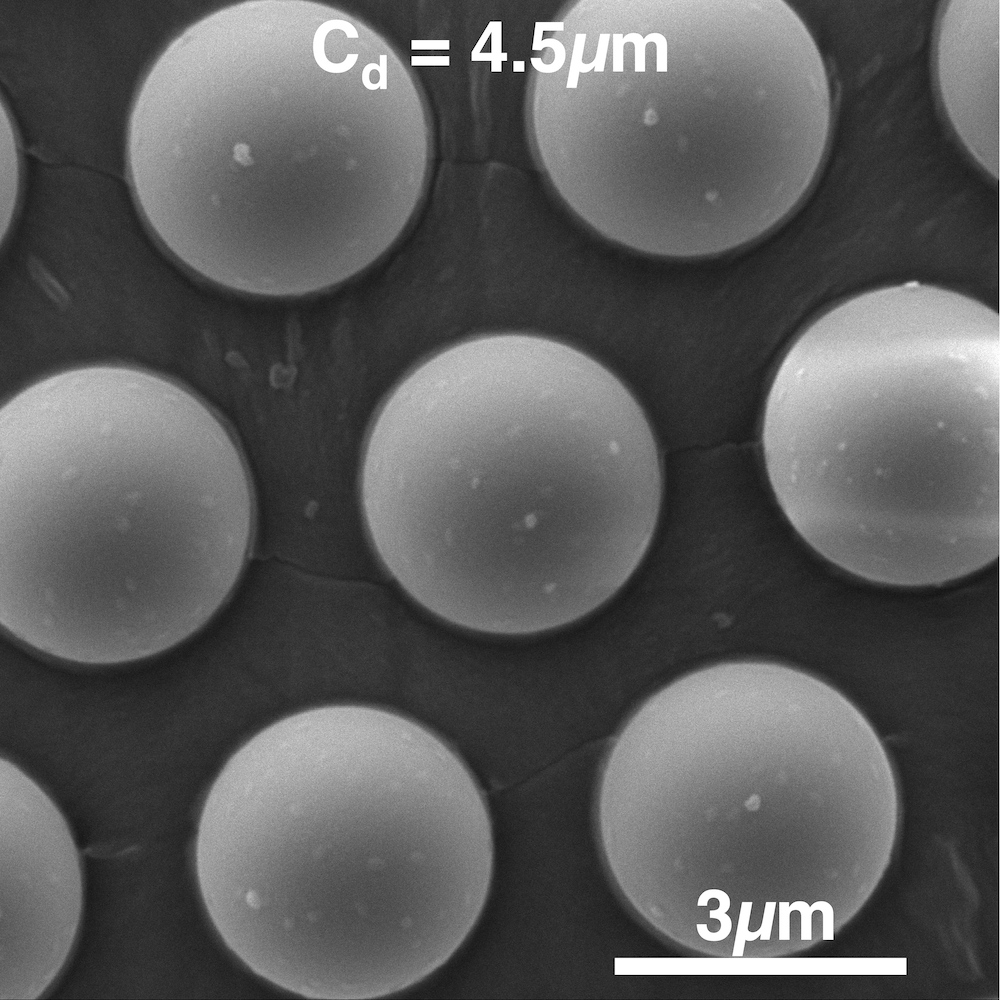}
        \caption{}\label{fig:cd45}
    \end{subfigure}

    \caption{SEM image of substrates with particles occupying the grooves at $c_d$: 
    (a) \qty{3.5}{\um}, 
    (b) \qty{4.0}{\um}, and 
    (c) \qty{4.5}{\um}.}\label{fig:cd}
\end{figure}

Center-to-center spacing $c_d$ is varied as shown in Figure~\ref{fig:cd}: $c_d\sim$~\qtylist[list-units = single]{3.5;4.0;4.5}{\um}. These results show that our method is highly robust for assembling a monolayer of microparticles with a desired lattice spacing.  

\section{Area calculation}\label{appendix:area}
\begin{figure}[H]
    \centering
    \begin{subfigure}[c]{0.6\textwidth}
        \includegraphics[width=\textwidth]{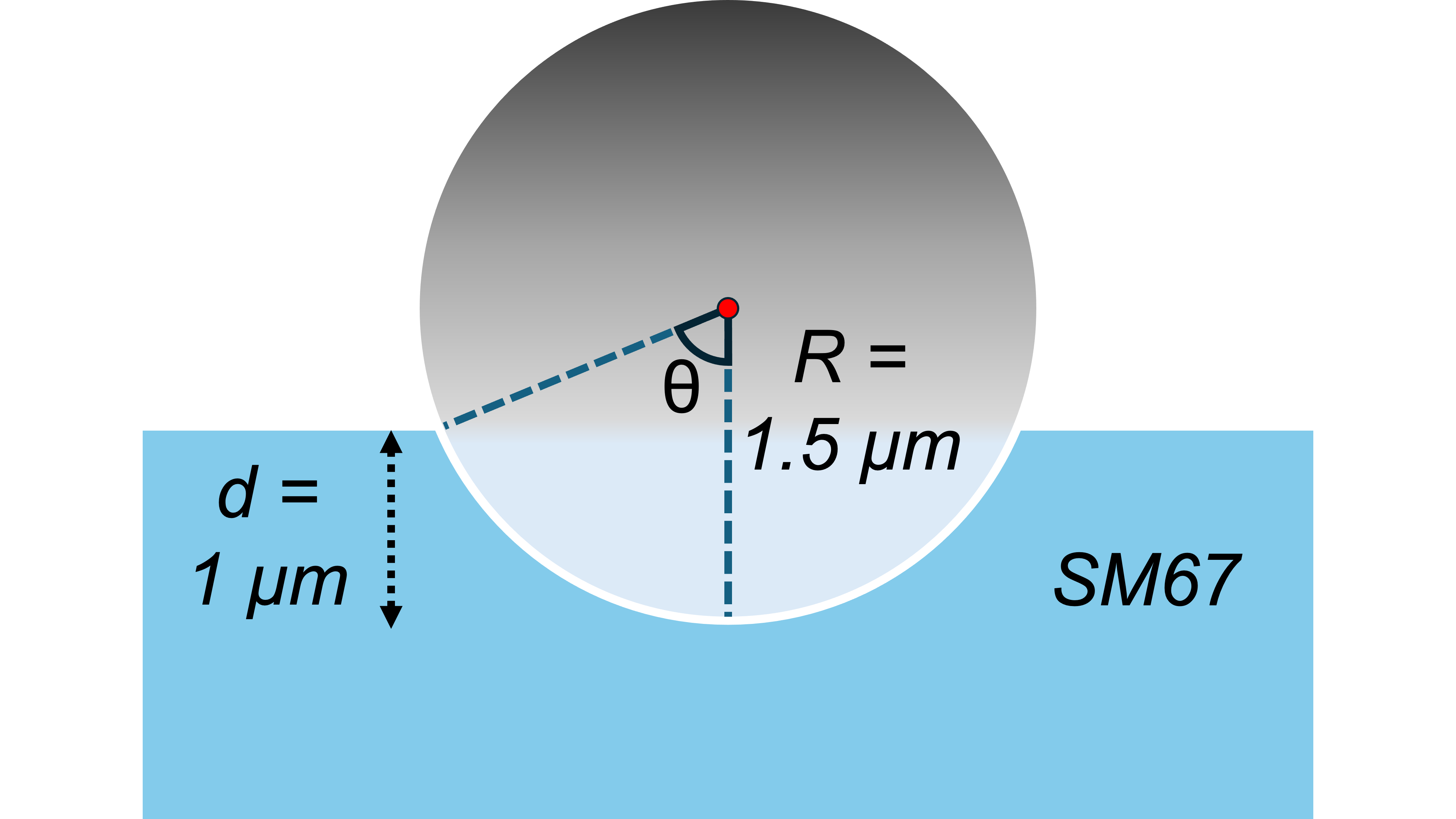}
    \end{subfigure}

    \caption{Schematic of microparticle embedded in a pattered substrate with groove depth \qty{1}{\um}, named SM67.}\label{fig:schematic_SM67_area}
\end{figure}

The surface area of a spherical cap (the portion of a sphere of radius \(R\) lying between the pole and polar angle \(\theta\)) follows directly from the surface measure in spherical coordinates.

The infinitesimal area element on the sphere is
\[
\mathrm{d}A=R^{2}\sin\theta'\,\mathrm{d}\theta'\,\mathrm{d}\phi.
\]
Integrating over azimuth \(\phi\in[0,2\pi]\) and polar angle \(\theta'\in[0,\theta]\) therefore gives
\[
A=\int_{0}^{2\pi}\int_{0}^{\theta}R^{2}\sin\theta'\,\mathrm{d}\theta'\,\mathrm{d}\phi=2\pi R^{2}\int_{0}^{\theta}\sin\theta'\,\mathrm{d}\theta'.
\]
The remaining integral follows:
\[
\int_{0}^{\theta}\sin\theta'\,\mathrm{d}\theta'=\bigl[-\cos\theta'\bigr]_{0}^{\theta}=1-\cos\theta,
\]
So the exact surface area is
\[
A=2\pi R^{2}(1-\cos\theta).
\]
Hence, the area available for deposition:
\[
A_{d}=4\pi R^{2} - A = 2\pi R^{2}(1+\cos\theta).
\]
and
\[
\cos\theta = \frac{R-d}{R}
\]
For SM67 (Figure~\ref{fig:schematic_SM67_area}), we have a groove depth of \qty{1.0}{\um} and R = \qty{1.5}{\um}, which corresponds to 66.67\% of the available area.

\section{Details about the diffusion coefficient of Janus particles}\label{appendix:diffusion}
\begin{figure}[htbp]
    \centering
    \begin{subfigure}[c]{0.6\textwidth}
        \includegraphics[width=\textwidth]{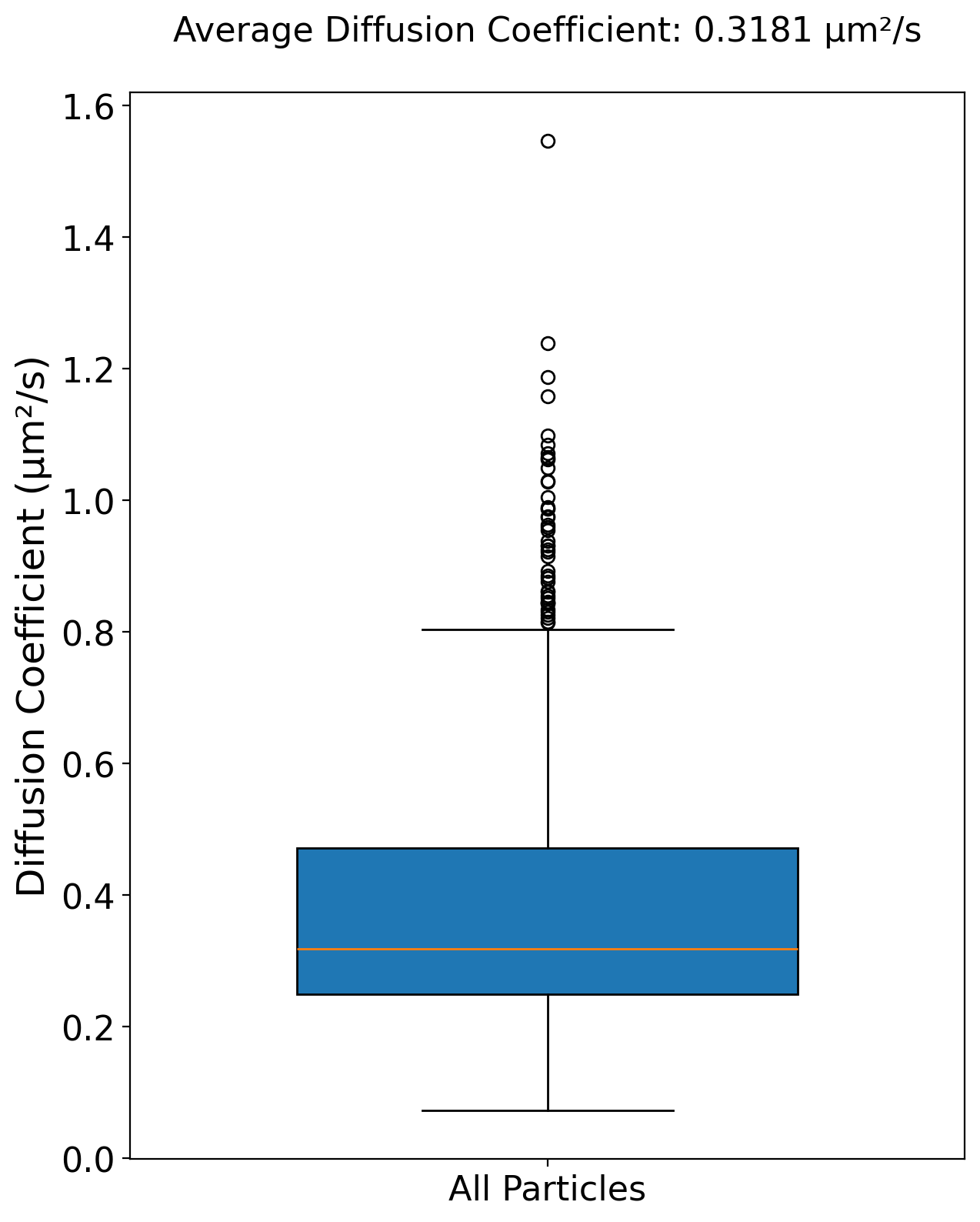}
    \end{subfigure}
    \caption{Distribution of diffusion coefficient experimentally derived from SM50 particles in 0\% \ce{H2O2}.}\label{fig:diffusion_boxplot}
\end{figure}

We calculated the diffusion coefficient of Janus Pt/silica particles prepared with CS substrates in ultrapure water. The same experimental and analysis method as discussed in the Experimental section were used, but without adding any \ce{H2O2} The mean square displacement was calculated from the analyzed trajectories of the particles, and MSDs were fitted with standard $4D_t \tau $ and the resulting $D_t$ value was \qty{0.3181}{\um^{2}/s} (Figure~\ref{fig:diffusion_boxplot}). This resulting $D_t$ value was used as a fitting parameter in the actual velocity calculations for the violin plot. The resulting $D_t$ was much higher than the theoretical value used for bare silica particles using Stokes-Einstein's relation: 
\[
D_t = K_bT/6\pi \eta r = 0.1455~\mu m^{2}/s.
\]

\section{FIB-SEM images}\label{appendix:fib_tagged}
\begin{figure}[H]
    \centering
    \begin{subfigure}[b]{0.32\textwidth}
        \includegraphics[width=\textwidth]{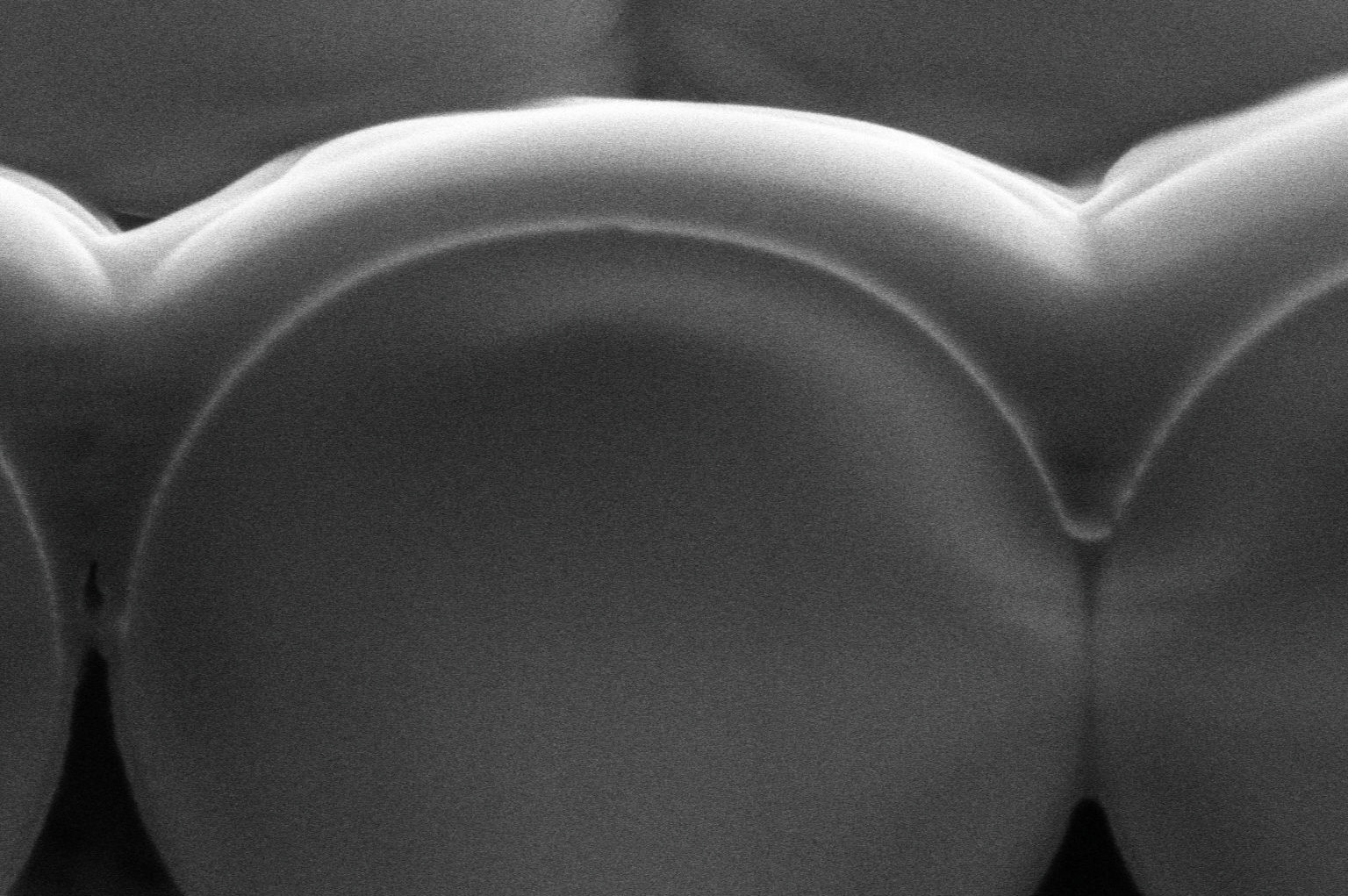}
        \includegraphics[width=\textwidth]{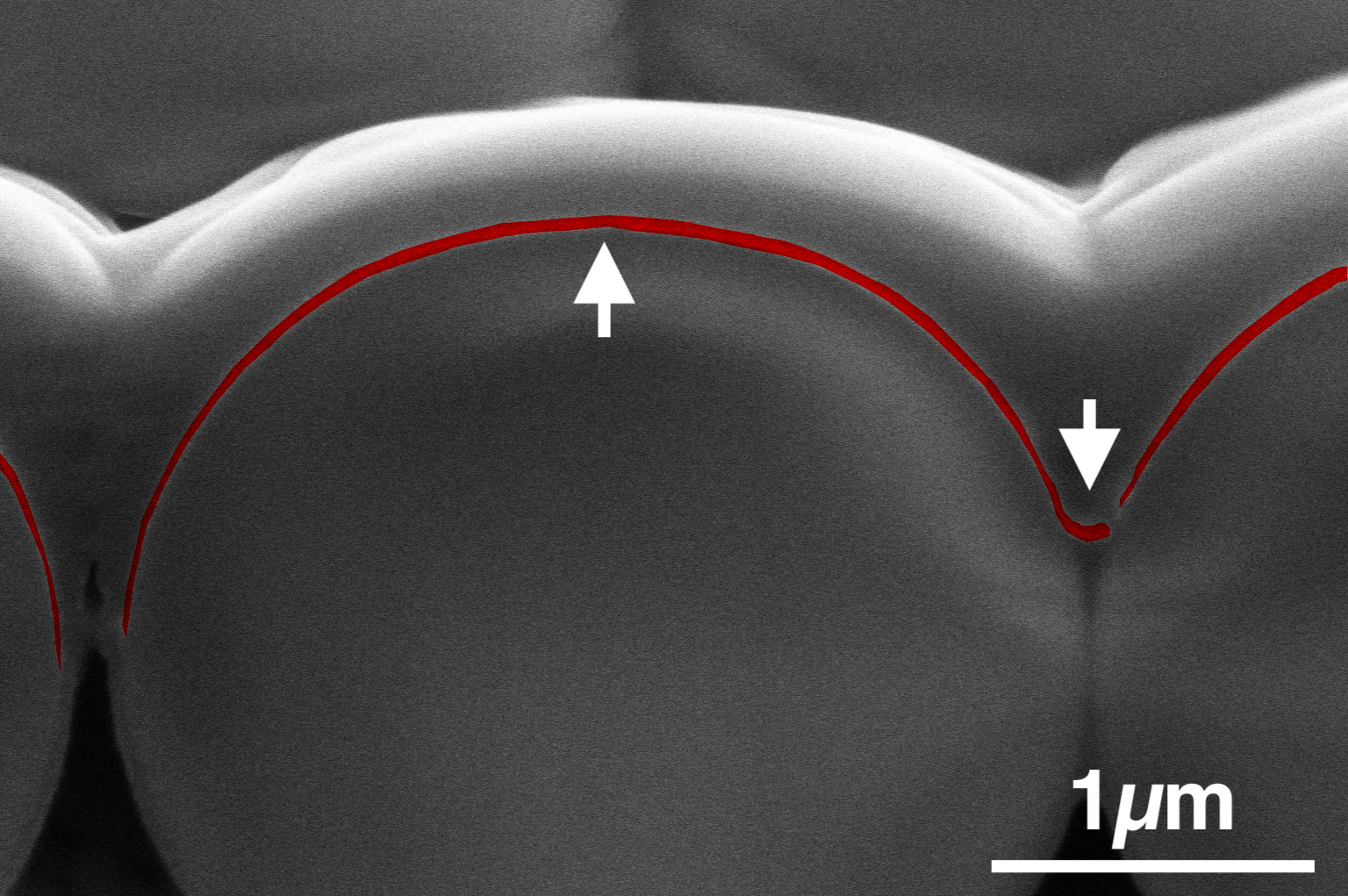}
        \caption{CM}\label{fig:fib_closed_original}
    \end{subfigure}
    \hfill
    \begin{subfigure}[b]{0.32\textwidth}
        \includegraphics[width=\textwidth]{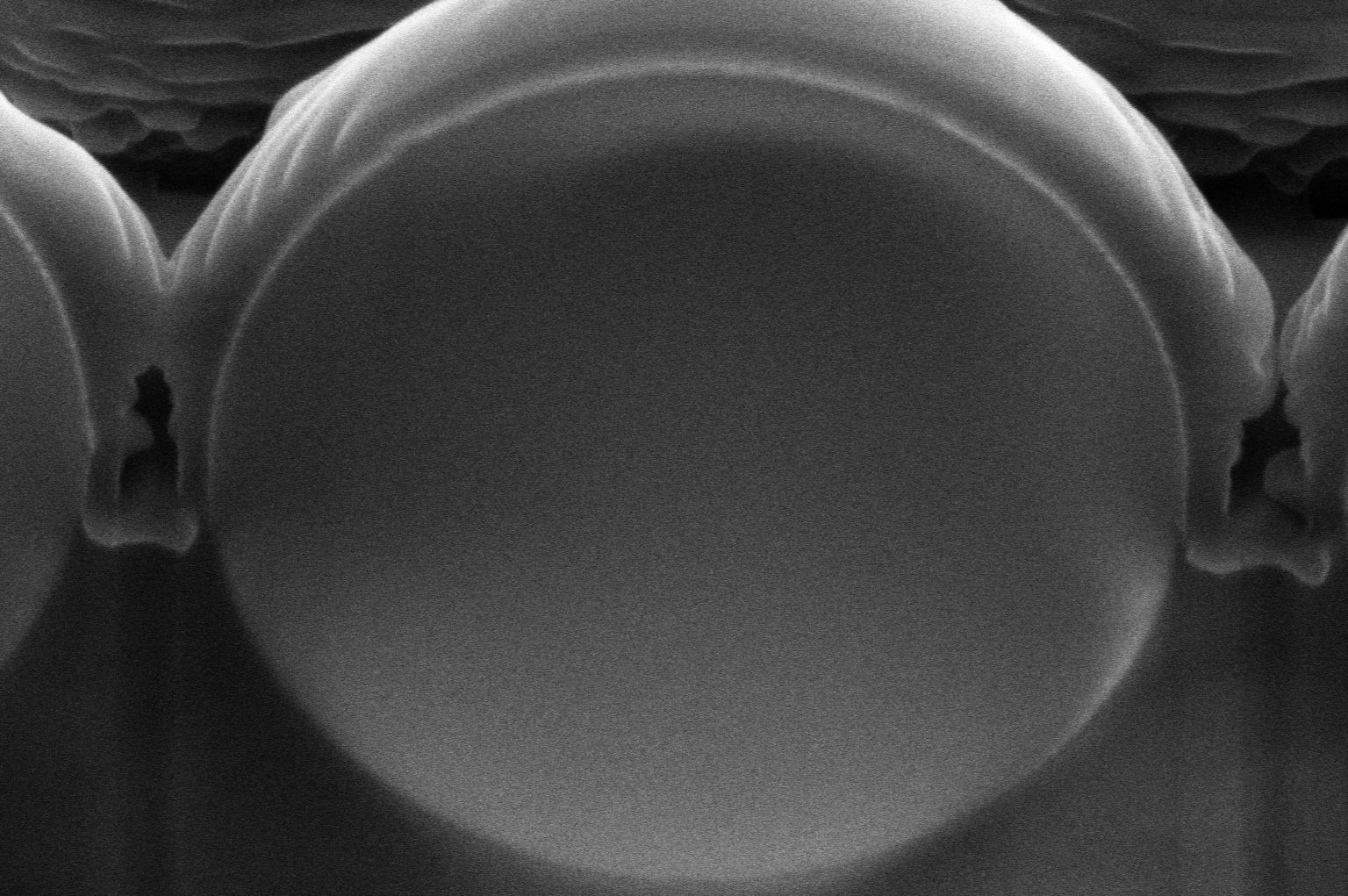}
        \includegraphics[width=\textwidth]{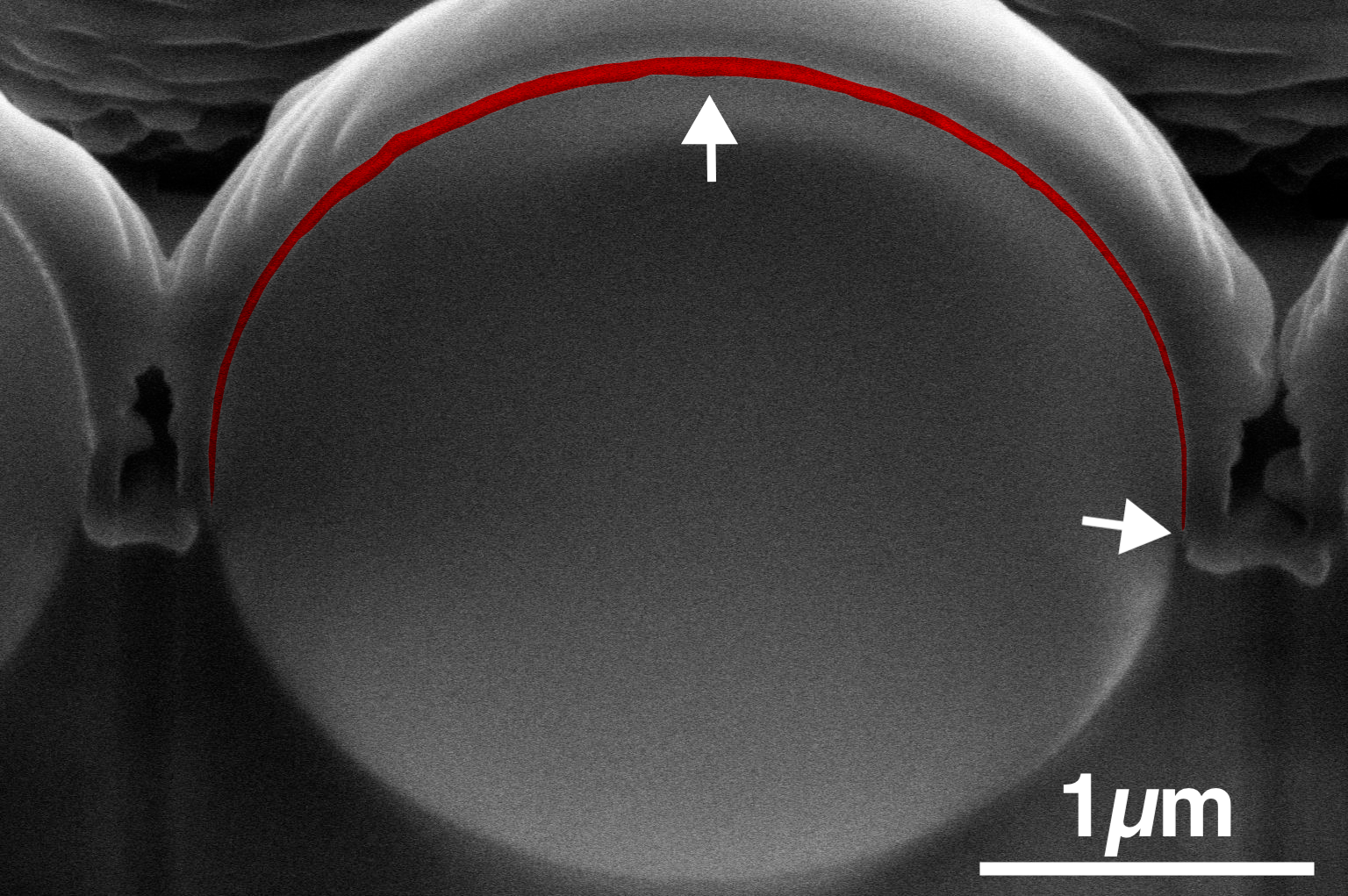}
        \caption{SM50}\label{fig:fib_sm50_original}
    \end{subfigure}
    \hfill
    \begin{subfigure}[b]{0.32\textwidth}
        \includegraphics[width=\textwidth]{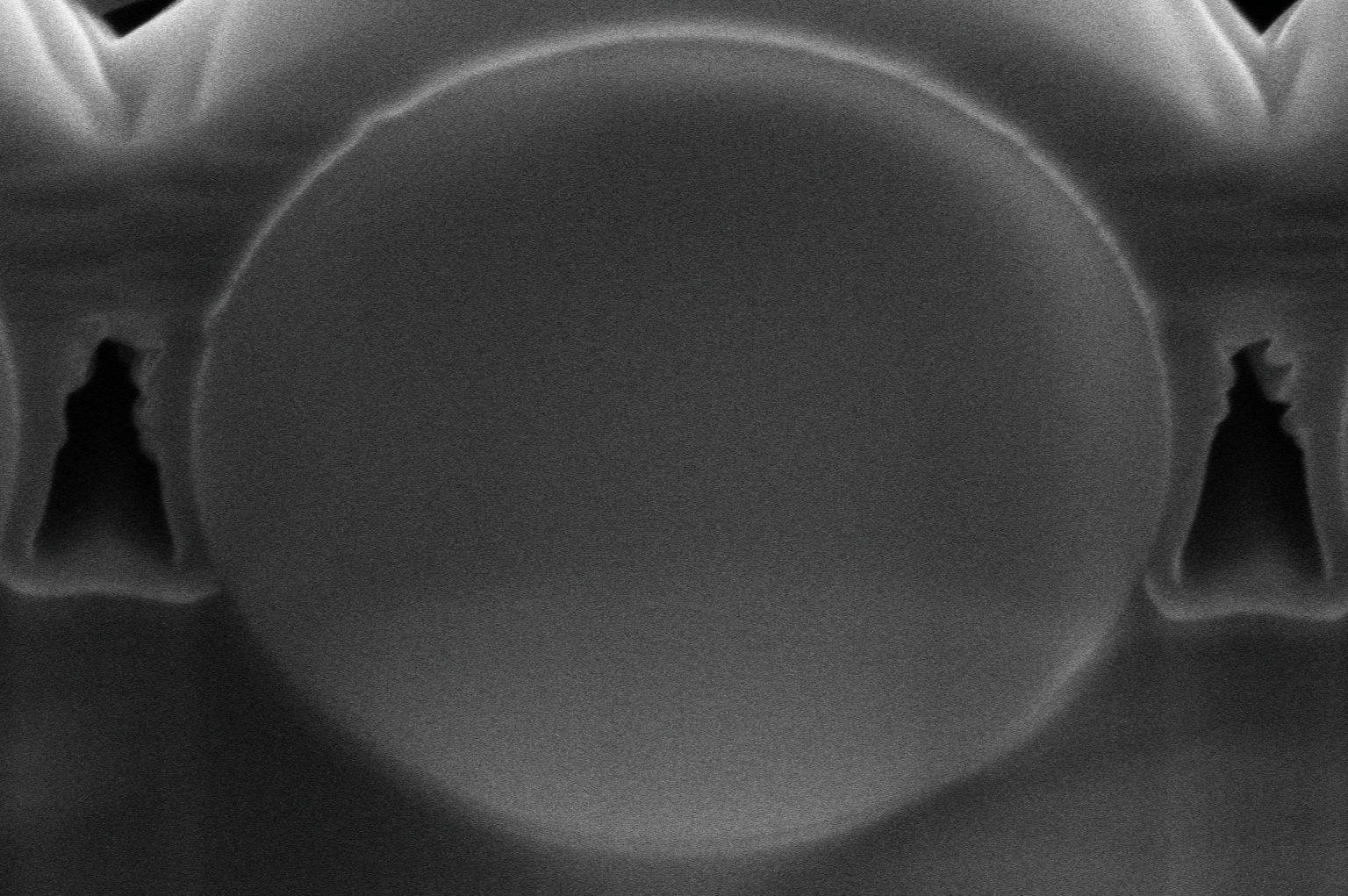}
        \includegraphics[width=\textwidth]{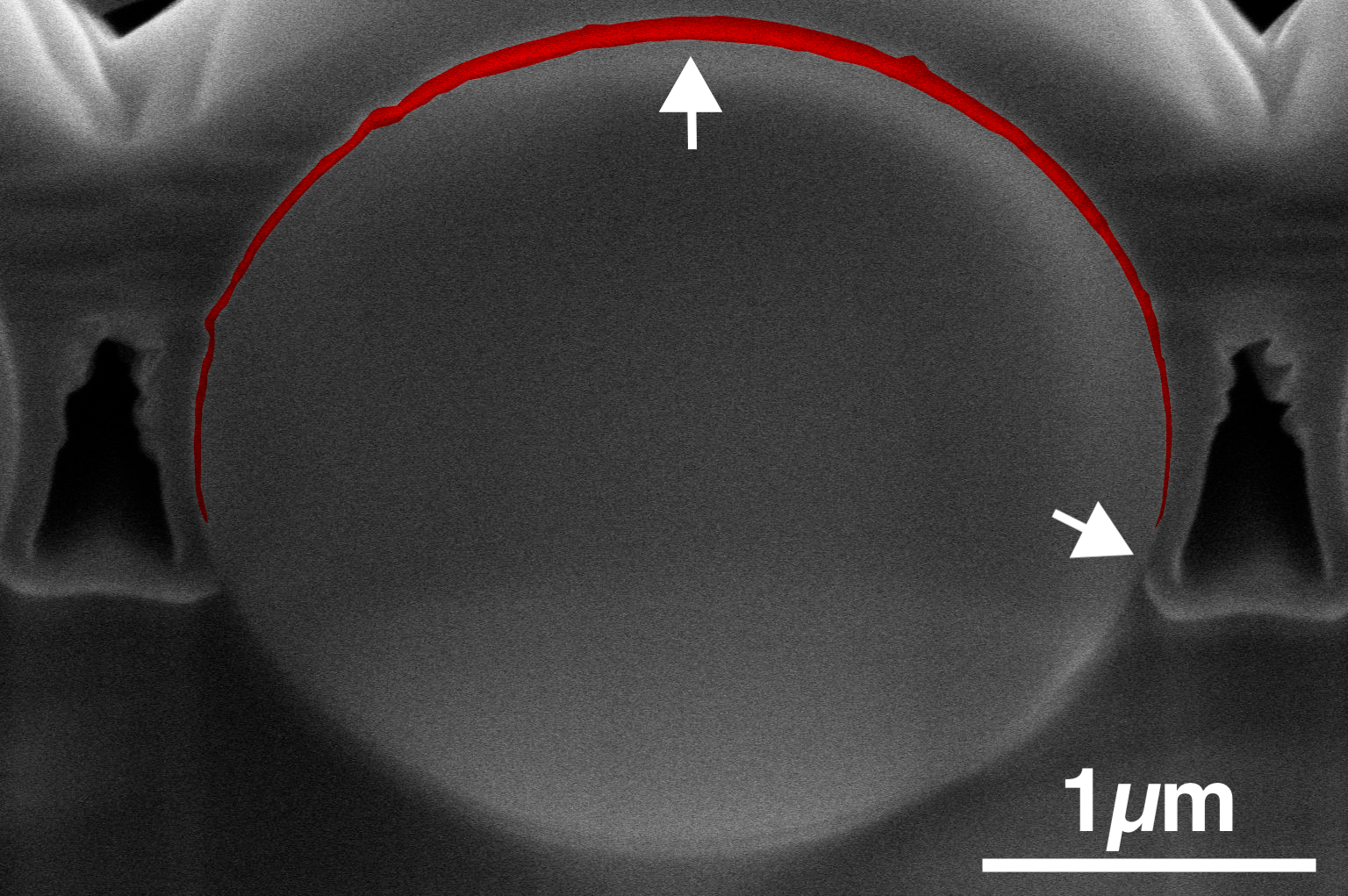}
        \caption{SM67}\label{fig:fib_sm67_original}
    \end{subfigure}

    \caption{Particle cross-section obtained using Focused Ion Beam (FIB) milling for particles in: 
    (a) closed packing monolayers CM, 
    (b) SM50 substrate, and 
    (c) SM67 substrate. 
 In each image, the \ce{Pt} is tagged in red and the corresponding original SEM is stacked above. The arrows are for guidance of \ce{Pt} layer.}
    \label{fig:fib_appendix}
\end{figure}

\bibliographystyle{elsarticle-num}
\bibliography{paper.bib}
\end{document}